%% file: VarSel.tex
\documentclass[12pt, a4paper]{article}

\usepackage{amsfonts}
\usepackage{amsmath, amsthm, amssymb}
\usepackage[round]{natbib}
\usepackage{dcolumn}
\usepackage{enumerate}
\usepackage{hhline}
\usepackage{dsfont}
\usepackage{textcomp}
\usepackage{lineno}
\usepackage[nolists, nomarkers]{endfloat}
\usepackage{graphicx}
\usepackage{color}
\usepackage[usenames,dvipsnames]{xcolor}
\usepackage{rotating}
\usepackage{hyperref}
\usepackage{multirow}
\usepackage{shortvrb}
\usepackage{xr}
\usepackage{rotating}
\usepackage{paralist}

\usepackage{floatrow}
\usepackage{float}
\usepackage[percent]{overpic}
\usepackage{subfig}
\usepackage[]{caption}
\usepackage{multicol}
\usepackage{gensymb}

\usepackage{fullpage}
\usepackage{pdflscape}
\usepackage{subfig}
\usepackage{bigstrut}
\usepackage{amsmath}
\usepackage{caption}

\usepackage{longtable}

\usepackage{environ}

\makeatletter
\NewEnviron{Lalign}{\tagsleft@true\begin{flalign}\hspace{3em}\BODY\end{flalign}}
\makeatother

\renewcommand{\baselinestretch}{1.2}

\newcounter{myremark}

\DeclareMathOperator{\BPD}{BP}
\renewenvironment{itemize}[1]{\begin{compactitem}#1}{\end{compactitem}}
\renewenvironment{enumerate}[1]{\begin{compactenum}#1}{\end{compactenum}}

\input{defs.tex}

\input{alm.tex}

\usepackage{titlesec}
\titlespacing*{\section}{0pt}{0.5\baselineskip}{0.5\baselineskip}
\titlespacing*{\subsection}{0pt}{0.5\baselineskip}{0.5\baselineskip}
\titlespacing*{\subsubsection}{0pt}{0.5\baselineskip}{0.3\baselineskip}

\newlength{\bibitemsep}
\newlength{\bibparskip}
\let\oldthebibliography\thebibliography
\renewcommand\thebibliography[1]{%
  \oldthebibliography{#1}%
  \setlength{\parskip}{\bibitemsep}%
  \setlength{\itemsep}{\bibparskip}%
}

\begin{document}
\setdefaultleftmargin{3.5mm}{3mm}{3mm}{3mm}{3mm}{3mm}
\title{Bayesian Effect Selection in Structured Additive Distributional Regression  Models}
\author{Nadja Klein$^{1\mbox{}^\star}$, Manuel Carlan$^2$, Thomas Kneib$^2$,\\\vspace{0.1cm} Stefan Lang$^3$ and Helga Wagner$^4$\\
 \normalsize $^1$Humboldt University of Berlin, $^2$Georg-August-Universit\"{a}t G\"{o}ttingen,\\\vspace{-0.2cm} \normalsize$^3$Universit\"{a}t Innsbruck, $^4$Johannes-Kepler-Universit\"{a}t Linz
 }
\date{ }

\maketitle

\begin{abstract}\footnotesize
\noindent
We propose a novel spike and slab prior specification with scaled beta prime marginals for the importance
parameters of regression coefficients to allow for
general effect selection within the class of structured additive distributional regression. This
enables us to model effects on all distributional parameters for arbitrary parametric distributions, and to consider various effect types such as non-linear or spatial effects as well
as hierarchical regression structures. Our spike and slab prior relies on a
parameter expansion that separates blocks of regression coefficients into overall scalar importance parameters and vectors of standardised coefficients. Hence, we can work with a scalar quantity for effect selection instead of a
possibly high-dimensional effect vector, which yields improved shrinkage and sampling performance compared
to the classical normal-inverse-gamma prior. We investigate the propriety of the posterior,
show that the  prior yields desirable shrinkage properties, propose a way of
eliciting prior parameters and provide efficient Markov Chain Monte Carlo sampling. Using both
simulated and three  large-scale data sets, we show that our approach is applicable for data
with a potentially large number of covariates, multilevel predictors accounting for hierarchically
nested data and  non-standard response distributions, such as bivariate normal or zero-inflated
Poisson.
\end{abstract}


\vfill
\noindent
{\small $\mbox{}^\star$  Correspondence should be directed to Prof.~Dr.~Nadja Klein at Humboldt University of Berlin, Spandauer Str.~1, 10178 Berlin. Email: nadja.klein@hu-berlin.de. The work of Manuel Carlan was supported by the German Research Foundation (DFG) via the research training group 1644 ``Scaling Problems in Statistics''. Thomas Kneib received financial support from the German Research Foundation (DFG) within the research project KN 922/9-1. Nadja Klein gratefully acknowledges funding by the Alexander von Humboldt Foundation.}

\newpage

\setlength{\abovedisplayskip}{0.15cm}
\setlength{\belowdisplayskip}{0.15cm}

\section{Introduction}\label{sec:Int}

The flexibility of modern regression methodology is both a blessing and a curse for applied researchers and statisticians alike since, on the one hand, added flexibility enables potentially more realistic models approximating the true data generating process but, on the other hand, poses additional challenges in the model building and model checking process. In this paper, we  consider structured additive distributional regression models \citep{RigSta2005,KleKneLanSoh2015} that combine additive predictors consisting of various types of regression effects, e.g.~non-linear effects of continuous covariates, spatial effects or random effects \citep{KamWan2003,RupWanCar2003,Woo2017} with the possibility to model all parameters of the response distribution (e.g.~location, scale or shape parameters) in terms of covariates in a distributional regression approach. As a consequence, an analyst is faced with the challenge of not only choosing an appropriate response distribution, \citep[a task that we will not consider in this paper since both graphical tools for model checking as well as selection criteria are well developed, see for example][]{KleKneLanSoh2015} but also with determining the most appropriate subset of covariates along with their exact modelling alternative for multiple regression predictors.

As an example, in one of our empirical illustrations on childhood undernutrition in Nigeria with more than 20,000 observations, we analyse a bivariate response variable $(y_1,y_2)'$ consisting of two scores for chronic and acute undernutrition. A previous study \citep{KleKneKlaLan2015} suggests a bivariate normal model in which not only the marginal expectations but also the marginal scale parameters and the correlation parameter depend on covariates. This leads  to a distributional regression model with five parameters $\mu_1,\mu_2,\sigma_1,\sigma_2,\rho$. In a full model, all of these parameters could be related to a predictor $\eta_{ik}$ of the form
\begin{align*}
\eta_{ik} =& \mathbf{\xvec}_i' \boldsymbol{\beta_{k}} + f_{1,k} ({\mathit{cage}}) + f_{2,k }({\mathit{mage}}) + f_{3,k}(\mathit{mbmi})+ f_{\mathit{spat},k} (\mathit{region}),\quad k=1,\ldots,5,\quad i=1,\ldots,n,
\end{align*}
where $i=1,\ldots,n$ denotes the observation index, $k$ refers to the five distributional parameters, $\xvec_i$ contains 13 binary covariates (and an intercept term) with regression coefficients $\betavec_k$, $f_{j,k}(\cdot)$, $j=1,2,3$, are non-linear smooth functions of age of child ($\mathit{cage}$), mother's age ($\mathit{mage}$) and mother's body mass index ($\mathit{mbmi}$), and $f_{\mathit{spat},k}$ are spatial effects based on regional information in the data. While effect selection (deciding which of the different effects should be included in the model) via a full search in the model space would already be challenging in a mean regression framework with only one single predictor, full effect selection in a distributional regression setting with multiple predictors is typically computationally prohibitive. This is even more the case when one is interested in deciding whether the effect of a continuous covariate shall be included in a linear or non-linear form or whether it could be excluded completely from the model. In this paper, we address these challenges and develop a novel spike and slab prior structure that enables Bayesian effect selection within structured additive distributional regression models.

While there has been extensive interest in spike and slab priors for Bayesian variable selection (i.e.~the selection of effects in models with purely linear predictors) or function selection (selection of non-linear effects of continuous covariates) in previous years \citep[see for example][for reviews]{ClyGeo2004,HarSil2009}, most research has been restricted to additive mean regression with Gaussian errors, distributions from the exponential family or survival models but also in the context of group variable selection~\citep{VeeMalMan2014,XuGho2015}. Furthermore, most approaches restrict the predictor specification to include either only linear effects or only non-linear effects of continuous covariates but do not enable the consideration of more complex effect types such as spatial effects or the decomposition of non-linear effects in linear and non-linear components. 

Classical Bayesian variable selection approaches for linear models based on spike and slab priors include for example \citet{MitBea1988}, or \citet{GeoMcC1997}. \citet{SmiKoh1996} utilise these approaches for function selection in nonparametric regression with Gaussian responses by assigning the variable selection priors to individual basis functions. Approaches that move beyond the framework of Gaussian models but pertain the purely linear predictor structure comprise the approaches of \citet{RosRub2017} who propose a Bayesian variable selection approach that allows for skewness and thicker tails compared to the Gaussian distribution, \citet{Wangetal2017} who consider variable selection after transforming the response, and \citet{ChuDun2009,KunDun2014} who propose non-parametric models where in the former proposal the mean and shape learn the effect of covariates, while the latter assumes symmetric residuals. In all these approaches however, the spike and slab prior is directly imposed on the scalar regression coefficients. In contrast, \citet{IshRao2005} consider a hierarchical specification where the spike and slab structure is not imposed directly on the regression coefficients but, on a higher level of the hierarchy, on their prior variances. This approach also allows to consider situations where selection should take place on blocks of regression coefficients representing for example the coefficients of a basis expansion in nonparametric regression. This leads to function selection approaches for additive models, also considered in  \citet{YauKohWoo2003,CotKohNot2008,ReiStoBon2009}, who combine a spike with point mass at zero with a slab that has support only on the positive real numbers. In contrast, \cite{ZhuVanCox2010} specify both spike and slab as normal distributions (with very different variance components) and \citet{PanSmi2008b} assign a multivariate prior with spike at the origin and normal slab directly to the whole vector of basis coefficients. In either case, one typically observes poor mixing unless sampling from marginalized full conditionals which are only available in closed form for Gaussian models \citep{YauKohWoo2003,ReiStoBon2009,PanSmi2008b} or models that have a latent Gaussian representation such as the probit model \citep{ZhuVanCox2010}. \cite{CotKohNot2008} address function selection in double exponential regression models, where both the mean and the dispersion parameter are linked to an additive predictor which comprises linear and non-linear effects. The model space is restricted, since functional effects may enter the model only if the corresponding linear effect is included in the model. 

Our proposal is inspired by the approach of \citet{schfah11} that introduces effect selection in generalized additive models for simple exponential family regression and with only one mean-related additive predictor. As \citet{schfah11}, we rely on a redundant parameter expansion of the vector of the basis coefficients as originally proposed in \citet{GelGykHuaBos2008}, and which allows us to expand the vector of basis coefficients in an importance parameter shared by all basis coefficients on the one hand and standardised basis coefficients on the other hand. Effect selection is then performed by assigning a spike and slab prior to the squared importance parameter. More precisely, our paper makes the following important contributions:
\begin{itemize}
 \item We integrate effect selection based on spike and slab priors in the structured additive distributional regression framework such that selection of general effect types is no longer restricted to mean regression models with responses from simple exponential families.
 \item The parameter vectors representing the additive effect components in a structured additive predictor are typically assigned partially improper multivariate normal priors. Instead of explicitly reparameterising the vector of basis coefficients to enable the specification of proper priors as in \citet{schfah11}, we implicitly remove the partial impropriety by adding a corresponding constraint to the prior distribution. As a consequence, we can retain sparse matrix structures for speeding up computations and show empirically that this has beneficial impact on the mixing behaviour of the MCMC simulations. In particular, when the vector of regression coefficients is large, we do not observe the strong dependence on the dimensionality of the basis coefficient vector identified in \citet{schfah11}. This enables us to also include effects of considerable dimension such as spatial effects to truly exploit the benefits of effect selection over function selection and even allows us to further extend the model to hierarchical specifications of the predictors~\citep{LanUmlWecHarKne2012}.
 \item Formulating the spike and slab prior for the squared importance parameter in the redundant parameterisation yields scaled beta prime marginals which have favourable shrinkage properties~\citep{PerPerRam2017}. We study these properties in detail and provide corresponding theoretical results for our prior structure including conditions for the propriety of the posterior.
  \item We develop rules for eliciting the hyperparameters of the spike and slab prior based on simple scaling criteria that are easily accessible to applied researchers. Based on the elicited parameters, we find that our new prior structure has similarly favourable shrinkage properties as the approach by \citet{schfah11}, while it avoids to arbitrarily fix the hyperparameters.
\end{itemize}
The rest of this paper is structured as follows: Section~\ref{sec:BayVarSelDisReg} summarises the specification of our novel spike and slab prior for effect selection in distributional regression. Properties of the prior, including prior elicitation, shrinkage properties and propriety of the posterior are discussed in Section~\ref{sec:NBPSS:properties}. Section~\ref{sec:Inf} contains details on posterior estimation via Markov chain Monte Carlo simulations and points to software and implementation. Sections~\ref{sec:sims} and \ref{sec:App} evaluate the performance of our approach in simulations and three diverse applications. In Section~\ref{sec:SumDis} we conclude.
\section{Bayesian Effect Selection in Distributional Regression}\label{sec:BayVarSelDisReg}
\subsection{Observation Model}\label{subsec:ModSep}
\subsubsection{Distributional Regression}
Our approach to Bayesian effect selection based on spike and slab priors is developed for the general class of (multivariate) Bayesian structured additive distributional regression \citep{KleKneLanSoh2015}. Let $(\yvec_i, \nuvec_i)$, $i=1,\ldots,n$ denote $n$ independent observations on the (not necessarily scalar) response variable $\yvec$ and covariates $\nuvec$. We then assume that the conditional distribution of $\yvec_i$ given $\nuvec_i$ is specified in terms of a $K$-parametric distribution with density
\[
 p(\yvec_i|\vartheta_{i1},\ldots,\vartheta_{iK}), \tag{M1}\label{eq:M1}
\]
where $\varthetavec_i=(\vartheta_{i1},\ldots,\vartheta_{iK})'$ is a collection of $K$ scalar distributional parameters $\vartheta_{ik}$, $k=1,\ldots,K$, which depend on $\nuvec_i$. Compared to mean regression models where $p(\cdot)$ is usually assumed to belong to the exponential family and where $K-1$ parameters are treated as fixed or nuisance parameters, in distributional regression each of the distributional parameters is linked to a structured additive predictor $\eta_{ik}$ via a suitable one-to-one transformation $h_k$, i.e.~$h_k(\eta_{ik})=\vartheta_{ik}$ and $\eta_{ik} = h^{-1}_k(\vartheta_{ik})$.
\subsubsection{Structured Additive Predictors}\label{subsec:pred}
The predictors themselves are specified as
\[\tag{M2}\label{eq:M2}
 \eta_{ik} = \eta_{ik}^{\mathrm{in}} + \eta_{ik}^{\mathrm{sel}} =  \sum_{l=1}^{L_k}f_{l,k}^{\mathrm{in}}(\nuvec_i) + \sum_{j=1}^{J_k}f_{j,k}^{\mathrm{sel}}(\nuvec_i),
\]
where the effects $f_{j,k}^{\mathrm{sel}}(\nuvec_i)$ represent various types of flexible functions depending on (different subsets of) the covariate vector $\nuvec_i$ that are to be selected via spike and slab priors, while $\eta_{ik}^{\mathrm{in}}$ represents a second additive predictor consisting of all effects $f_{l,k}^{\mathrm{in}}(\nuvec_i)$ that are \emph{not} under selection. The separation into two subsets of effects allows us to include specific covariate effects mandatorily in the model (e.g.~based on prior knowledge or since these represent confounding effects that have to be included in the model in any case). In the following, we will only discuss the specification of priors for the effects under selection in detail since the effects $\eta_{ik}^\mathrm{in}$ can be handled exactly as in distributional regression models without effect selection, but we will use the differentiation later in Section~\ref{propriety} for deriving sufficient conditions for the propriety of the posterior.

Dropping the parameter index $k$, the function index $j$ and the superscript ${\mathit{sel}}$ in the rest of this section for notational simplicity, we assume that each effect $f(\nuvec_i)$ can be approximated by a linear combination of basis functions such that
\[\tag{M3}\label{eq:genericeffect}
 f(\nuvec_i) = \tau\sum_{d=1}^{D}\tilde\beta_{d}B_{d}(\nuvec_i),
\]
where $B_d(\nuvec_i)$, $d=1,\ldots,D$ are the basis functions, $\betatildevec=(\tilde\beta_1,\ldots,\tilde\beta_D)'$ is the vector of (standardised) basis coefficients and $\tau$ is an importance parameter. Due to the linear basis representation, the vector of function evaluations $\fvec=(f(\nuvec_1),\ldots,f(\nuvec_n))'$ can be written as $\fvec=\tau\mB\betatildevec$ where $\mB$ is the ($n\times D$) design matrix arising from the evaluation of the basis functions $B_d(\nuvec_i)$, $d=1,\ldots,D$ at the observed covariate values $\nuvec_1,\ldots,\nuvec_n$.

Note that the parameterisation in (\ref{eq:genericeffect}) is equivalent to the standard specification in structured additive regression
\[ \tag{M3$^*$}\label{eq:geneffa}
 f(\nuvec_i) = \sum_{d=1}^{D}\beta_{d}B_{d}(\nuvec_i),
\]
but redundant as only the product $\betavec=\tau\betatildevec$ is identified.
 However, the importance parameter $\tau$ allows us to remove effects from the predictor for $\tau=0$ while effects are considered to be of high importance if $\tau$ is large in absolute terms. We will place a spike and slab prior on the squared importance parameter $\tau$ to achieve effect selection.

\subsection{The Normal Beta Prime Spike and Slab Prior}\label{subsec:NBPSS}
\subsubsection{Constraint Prior for Regression Coefficients}
Since for many specific types of effects the vector of basis coefficients $\betavec$ is of relatively high dimension, it is often useful to enforce specific properties such as smoothness or shrinkage. In a Bayesian formulation, this can be facilitated by assuming (partially improper) multivariate Gaussian priors
\[\tag{M4$^*$}\label{eq:M3a}	
 p(\betavec|\tau^2)\propto\exp\left(-\frac{1}{2\tau^2}\betavec'\mK\betavec\right)\mathds{1}\left\lbrack\mA\betavec=\nullvec\right\rbrack,
\]
where $\mK$ denotes the prior precision matrix implementing the desired properties, $\tau^2$ is a prior variance parameter and the indicator function $\mathds{1}[\mA\betavec= \nullvec]$ is included to enforce linear constraints on the regression coefficients via the constraint matrix $\mA$. The latter is typically used to remove identifiability problems from the additive predictor (e.g.~by centering the additive components of the predictor) but can also be used to remove the partial impropriety from the prior that comes from a potential rank deficiency of the precision matrix $\mK$ with $\rank(\mK)=\kappa\le D$.

We specify a prior of exactly the same structure on the vector of scaled basis coefficients $\betatildevec$,
\[\tag{M4}\label{eq:M3}	
 p(\betatildevec)\propto\exp\left(-\frac{1}{2}\betatildevec'\mK\betatildevec\right)\mathds{1}\left\lbrack\mA\betatildevec=\nullvec\right\rbrack
\]
and assume that the constraint matrix $\mA$ is chosen such that all rank-deficiencies in $\mK$ are effectively removed from the prior distribution. This can, for example, be achieved by setting
\[\tag{M5}\label{eq:M4}	
 \mA=\spa\left(\ker(\mK)\right),
\]
where $\ker(\mK)$ denotes the null space of $\mK$ and $\spa\left(\ker(\mK)\right)$ is a representation of the corresponding basis. This specification effectively restricts the parameter vector $\betatildevec$ to a lower dimensional space of dimension $\rank(\mK)$ and allows us to establish a decomposition of the effect $f(\nuvec)$ into a penalized and an unpenalized part, i.e.~ $f_{\unpen}(\nuvec)+f_{\pen}(\nuvec)$ where $f_{\unpen}(\nuvec)$ represents parts of the function corresponding to the null space of $\mK$ which are therefore not affected by the ``penalisation'' induced by $\mK$ while $f_{\pen}(\nuvec)$ represents the part of the total effect that is associated with the proper, informative prior part. Importantly, we can now put separate spike and slab priors on both parts of $f$. For instance, in case of penalized splines with second order random walk prior, the space of unpenalized functions contains the linear functions, while the penalized part contains nonlinear deviations from the former. Such a parameterization hence enables the decision whether a continuous covariate should be included purely nonlinearly, whether it is sufficient to assume a pure linear effect or whether the sum of a linear and a non-linear effect is needed. The resulting models are therefore both potentially more parsimonious and easier to interpret.

The specifications \eqref{eq:genericeffect}, \eqref{eq:M3} and \eqref{eq:geneffa}, \eqref{eq:M3a} seem to be equivalent to each other corresponding to rescaling the regression coefficients and the prior distribution as $\betavec=\tau\betatildevec$. However, this is only true if the prior distribution (\ref{eq:M3}) is indeed proper. To see this, assume that $\mK$ is rank deficient and a constant effect is not penalised by the prior precision matrix. In this case, the traditional formulation of structured additive regression models \eqref{eq:geneffa} implies a constant effect if $\tau^2$ approaches zero while the rescaled version \eqref{eq:genericeffect} implies an effect equal to zero since the complete function is multiplied by $\tau$.

Note, that both~\eqref{eq:M3a} and~\eqref{eq:M3} rely on the same precision matrix $\mK$ and hence the constraint matrix $\mA$ can be constructed independently of the parametrisation.
The traditional way is an explicit mixed model decomposition~\citep{FahKneLan2004,Woo2011} which is used by \cite{schfah11} to perform effect selection for mean regression models. As the mixed model representation yields a penalised component which is  $\betatildevec\sim\ND(\nullvec,\mI)$, this is effectively equivalent to considering our constraint prior by choosing the constraint matrix according to (\ref{eq:M4}) and by rescaling the individual entries in $\betatildevec$ with the eigenvalues of $\mK$ \citep[see][Sec.~3.2 for details]{RueHel2005}. However, the explicit mixed model representation used by \cite{schfah11} destroys the sparsity properties of the design matrices (such as band structures for B-splines) and causes full design matrices which in turn increases computation times. In order to keep the sparsity of the design matrices of functional effects (and hence to minimize computation time) we instead implicitly remove the improper part of $p(\betavec|\tau^2)$ by sampling $\betavec$ directly from the constrained posterior using~\eqref{eq:M3}.

\subsubsection{Normal Beta Prime Spike and Slab Prior on Squared Importance Parameter}
To achieve function selection in our model, we place a spike and slab prior specification on the squared importance parameter $\tau^2$. This hierarchical prior relies on a mixture of one prior concentrated close to zero such that it can effectively be thought of as representing zero (the spike component) and a more dispersed, mostly noninformative prior (the slab) and is specified via the hierarchy
\begin{equation}\tag{M6}\label{eq:M5}	
 \begin{aligned}
 \tau^2|\delta,\psi^2 & \sim\GaD\left(\frac{1}{2},\frac{1}{2r(\delta) \psi^2}\right) \\
 \delta|\omega   & \sim \BiD(1,\omega)\\
 \psi^2    & \sim \IGD(a,b)\\
 \omega    & \sim \BetaD(a_{0},b_{0})\\
 r(\delta) &=\begin{cases} r & \delta=0\\
                          1 & \delta=1
            \end{cases}
 \end{aligned}
\end{equation}
The scale parameter $\psi^2$ determines the prior expectation of $\tau^2$, which is $\psi^2$ for $\delta=1$ and $r \psi^2$ for $\delta=0$ with $r\ll1$ being a fixed small value and hence the indicator $\delta$ determines whether a specific effect $\betavec=\tau \betatildevec$ is included in the model ($\delta=1$) or excluded from the model ($\delta=0$). The parameter $\omega$ is the prior probability for an effect being included in the model and the remaining parameters $a$, $b$, $a_0$, $b_0$ and $r$ are hyperparameters of the spike and slab prior. We will discuss prior elicitation for these parameters in detail in Section~\ref{subsec:PriEli}.

Marginalising over $\psi^2$, both the spike and the slab component $p(\tau^2|\delta)$ are scaled beta prime distributions with shape parameters $1/2$ and $a$ and scale parameter $2r(\delta)b$~\citep{PerPerRam2017}. Therefore we call the hierarchical prior on $\betavec=\tau \betatildevec$ specified by (\ref{eq:M3})~--~(\ref{eq:M5}) the Normal Beta Prime Spike and Slab (NBPSS) prior, see Section~\ref{sec:NBPSS:properties} for a detailed discussion of the properties of the NBPSS prior. Equations (\ref{eq:M1}) to (\ref{eq:M5}) define our complete model specification for effect selection in structured additive distributional regression.

\subsection{Special Cases}

We briefly discuss some of the components of structured additive predictors used later in our empirical evaluations. These include
\begin{itemize}
\item {linear effects} with either flat, improper priors if these are not under selection or conditionally i.i.d.~Gaussian priors for linear effects under selection. The columns of the design matrix $\mB$ are then equal to the different covariates.
\item {non-linear effects} based on Bayesian P-splines~\citep{LanBre2004}, where random walk priors are used for the regression coefficients corresponding to $D$ different B-spline basis functions. The $i$-th row of $\mB$ then contains the basis functions $B_1(x_i),\ldots,B_D(x_i)$ evaluated at $x_i$. If not stated otherwise, we will use second order random walk priors and cubic B-splines with 20 inner knots resulting in $D=22$.
\item {spatial effects} for a discrete set of geographical regions modelled via Gaussian Markov random fields (GMRFs) with precision matrix given by an adjacency matrix encoding the neighbourhood relation between the regions~\citep{RueHel2005} and a design matrix with entries $(i,s)$ equal to one if observation $i$ is located in region $s$ and zero otherwise. We  consider the simplest form of GMRFs and define two regions as neighbours if they share common borders.
\item multilevel structured additive regression models as proposed by~\cite{LanUmlWecHarKne2012} that allow for hierarchical prior specifications for regression effects where each parameter vector may again be assigned an additive predictor, i.e.~the vector $\betavec$ is decomposed as $\betavec=\etavec+\varepsilonvec$ and the predictor $\etavec$ can itself  be of structured additive form. 
\end{itemize}

\section{Properties of the NBPSS prior}\label{sec:NBPSS:properties}
In the following, we discuss properties of the NBPSS prior hierarchy, including elicitation of hyperparameters, shrinkage properties and propriety of the posterior. For prior elicitation and shrinkage properties, the marginal distribution of $\betavec=\tau\betatildevec$ plays a crucial role. We will therefore start with deriving this marginal distribution.

\subsection{Marginal Distribution}\label{subsec:margprior}
The marginal prior for the squared importance parameter $\tau^2$  is given by the mixture
\begin{equation}\label{eq:mixdens_tau2}	
 p(\tau^2)= p(\tau^2|\delta=1) \dsP(\delta=1|a_0,b_0) +p(\tau^2|\delta=0) \dsP(\delta=0|a_0,b_0)
\end{equation}
of two scaled beta prime distributions $\BPD(1/2,a,2b)$ and $\BPD(1/2,a,2rb)$ with  mixture weight of the slab given by $\dsP(\delta=1|a_0,b_0)=a_0/(a_0+b_0)$. A modified version of the NBPSS prior can alternatively be derived by assuming a mixture of two scaled t distributions for the importance parameter $\tau=\pm \sqrt{\tau^2}$. Specifying this prior hierarchically, \mbox{the first equation in (\ref{eq:M5}) is replaced by}
$
 \tau|\delta,\psi^2 \sim\ND\left(0,r(\delta) \psi^2\right)
$
and as a consequence posterior sampling would no longer be possible with Gibbs steps as the corresponding conditional posterior would depend on the likelihood function.
Marginalising over $\psi^2$, $\delta$ and $\omega$, the prior $p(\tau)$ is a mixture of two scaled t-distributions with $2a$ degrees of freedom, location parameter 0, scale parameters $b/a$ and $rb/a$ and mixture weights $a_0/(a_0+b_0)$ and $b_0/(a_0+b_0)$, respectively. Thus, the prior on the (signed) importance parameter $\tau$ is closely linked to the NMIG prior used in~\citet{IshRao2005} when considering scalar regression coefficients $\beta$ that are conditionally normal given the inverse gamma distributed variance parameter $\tau^2$ (but with one level of hierarchy less) on the one hand, and, on the other hand to the peNMIG specification of~\citet{schfah11}.

The implied marginal distribution for $\betavec=\tau\betatildevec$ can now be derived as
\begin{equation}\label{eq:marg:prior}
\begin{aligned}
p(\betavec) &= \int_{-\infty}^\infty p(\tau)p_{\betatildevec}(\betatildevec/\tau)\frac{1}{|\tau|}d\tau,\\
\end{aligned}
\end{equation}
where $p_{\betatildevec}$ is given in equation (\ref{eq:M3}). However no analytical solution exists for this integral such that it has to be approximated numerically.
\subsection{Prior Elicitation}\label{subsec:PriEli}
In the following, we discuss prior elicitation for the NBPSS prior hyperparameters $a$, $b$, $a_0$, $b_0$ and $r$. More precisely, we argue that suitable default values can be suggested for $a$, $a_0$, and $b_0$ based on theoretical arguments while providing intuitive and user-friendly criteria for the elicitation of $b$ and $r$. In the literature, default values have often been suggested from simulation-based evidence \citep[e.g.~in][]{schfah11} but we prefer to determine $b$ and $r$ in a more transparent way. 

Theoretical properties of the scaled beta prime distribution have been discussed in~\citet{PerPerRam2017}. From this, it follows that for both spike and slab moments of order less than $a$ exist and the variance decreases with $a$. Furthermore, for small values of $a$, the spike and the slab component will overlap such that moves from $\delta=0$ to $\delta=1$ are possible. However to guarantee the existence of moments, $a$ should not be too small either. Fixing $a=5$ yielded overall a convincing mixing performance and we therefore use this value also in our real data examples.

For the prior inclusion parameter $\omega$ a sensible default is to use $a_0=b_0=1$ which corresponds to a flat prior on the unit interval. Of course, one can also choose fix values for $\omega$ in case strong prior knowledge on the prior inclusion probability of the size of the expected model is available. As the marginal prior inclusion probability is given by $\dsP(\delta=1|a_0,b_0)=a_0/(a_0+b_0)$, $a_0$ and $b_0$ can be chosen to reflect prior assumptions on the inclusion probability of effects.

For the elicitation of $b$ and $r$, we propose an approach inspired by the principled approaches of~\citet{SimRueMarRieSor2017} and \citet{KleKne2016}. More precisely, we consider marginal probability statements on the supremum norm $\sup_{\nuvec\in\calD}|f(\nuvec)|$ over a certain set of covariate values $\calD$ conditional on the status of the inclusion/exclusion parameter $\delta$. Given $\delta=1$ (inclusion of the effect), the marginal distribution of $f(\nuvec)$ does no longer depend on $r$, such that the parameter $b$ can be determined from
\begin{equation}\label{eq:cond:hyper2}
 \dsP\left(\left.\sup_{\nuvec\in\calD}|f(\nuvec)|\,\leq c\,\right|\,\delta=1\right) = \alpha,
\end{equation}
This is the probability that the supremum norm of an effect is smaller than a pre-specified level $c$ for all design points $\nuvec\in\calD$, such that $\alpha$ and $c$ should be small. Basically we formulate the prior such that it is unlikely that the supremum norm stays below a pre-specified level if it is indeed an informative effect that should be included. Both the level $c$ and the prior probability $\alpha$ have to be specified by the analyst according to her/his prior beliefs. To derive $r$, we proceed similarly but consider the probability
\begin{equation}\label{eq:cond:hyper1}
 \dsP\left(\left.\sup_{\nuvec\in\calD}|f(\nuvec)|\,\leq c\,\right|\,\delta=0\right) = 1-\alpha
\end{equation}
now conditioning on non-inclusion. Since in this case we would rather be interested in making the probability of not exceeding the threshold $c$ large, the probability is reversed to $1-\alpha$. Note that the absolute value of the effects can be taken without loss of generality due to the centring constraint of each function to ensure identifiability.

The basic idea of these two equations is that such prior statements can be much more easily elicited in applications, in particular in distributional regression where the application of response functions such as the exponential function or the logit transform induce default ranges of plausible effect sizes. Of course, the levels $c$ as well the probability levels $\alpha$ can be chosen to be distinct for the inclusion/exclusion criteria in (\ref{eq:cond:hyper2}) and (\ref{eq:cond:hyper1}) but we suppress this possibility notationally both for simplicity and since in most cases it seems plausible to choose the same parameter settings anyway.

To access the probabilities in (\ref{eq:cond:hyper2}) and (\ref{eq:cond:hyper1}), we have to derive the marginal distribution of $\sup |f(\nuvec)|$ which is not analytically accessible. For a single covariate value $\nuvec$, the function evaluation is given by $f(\nuvec)=\tau(B_1(\nuvec),\ldots,B_D(\nuvec))\betatildevec=\tau\bvec_{\nu}'\betatildevec=\bvec_{\nu}'\betavec$ and the marginal density is
\begin{equation*}
 p(\bvec_{\nu}'\betavec\,|\,\delta)=
 \int_0^\infty p(\bvec_{\nu}'\betavec|\tau^2)p(\tau^2\,|\,\delta)\mathrm{d}\tau^2
\end{equation*}
where $\bvec_{\nu}'\betavec|\tau^2 \sim\ND(0,\tau^2\bvec_{\nu}'\mK^{-}\bvec_{\nu})$ (with $\mK^{-}$ denoting the generalized inverse of $\mK$) and $p(\tau^2)$ is given in Equation (\ref{eq:mixdens_tau2}). Note that using the generalized inverse effectively removes the portion of $f(\nuvec)$ that corresponds to the null space of $\mK$ such that we take the constraint in (\ref{eq:M3}) into account. The integrals above are scalar integrals for each covariate $\nuvec$ which can be solved numerically. However, obtaining the supremum over a large set $\mathcal D$, numerical integration easily becomes computationally intractable. We hence determine the distribution of the supremum based on simulations from the hierarchical NBPSS prior.

In the Online Appendix~B, we show how to determine $r$ and $b$ independently of each other. For given design matrix $\mB=(\bvec_{\nu_1}',\ldots,\bvec_{\nu_n}')'$, precision matrix $\mK$, probability level $\alpha$ and threshold $c$, these can be computed for general functional effects using the R package \texttt{sdPrior}~\citep{sdPrior}.
\subsection{Shrinkage Properties}\label{subsec:ShrPro}
Regularisation and shrinkage properties of certain prior settings in regression specifications can be studied by considering the marginal distribution of the regression coefficients and/or functional effects. According to Section~\ref{subsec:margprior} the marginal densities have to be determined by numerical integration.
\subsubsection{Constraint Regions}
We compare the prior specified in \eqref{eq:M3}--\eqref{eq:M5} with a standard NMIG prior applied directly to the coefficients in $\betavec$ and the parameter expanded prior (peNMIG) of~\citet{schfah11}.
Figure~\ref{fig:marginaldensities1d} shows the univariate marginal log-densities where the most distinct difference is between the standard NMIG prior compared to peNMIG and NBPSS priors. While the standard NMIG prior resembles the shape of a normal distribution with a finite asymptote at zero, both parameter expanded priors feature a spike in zero. As we will show in the next section, this spike is indeed infinite such that advantageous selection behaviour is to be expected for the NBPSS prior.
Figure~\ref{fig:marginaldensities2d} supplements the univariate considerations by bivariate marginal log-densities. We differentiate between two situations: First, we consider two parameters that depend on the same value $\tau^2$, i.e.~parameters belonging to the same function $f(\nuvec)$, while in the second case we consider parameters depending on different importance parameters. This distinction is important since the standard NMIG prior always assumes independent components with separate hyperparameters. As a consequence, the peNMIG and NBPSS priors deviate from the standard situation in two ways: First by the parameter expansion itself and second by making the parameters depend on the same hyperparameter. To disentangle the effect of these two deviations, we rely on the separate presentations. We make the following important observations:
\begin{itemize}
 \item The NBPSS and peNMIG priors share the same qualitative behaviour while deviating considerably from the standard NMIG prior regardless of whether the case of shared or distinct $\tau^2$ is considered.
 \item The univariate marginal densities qualitatively resemble the ones of the original spike and slab prior of~\citet{MitBea1988} with tails that are heavy enough to induce a re-descending score function which ensures robustness of the Bayesian estimators (see also the next subsection).
 \item For the case of distinct parameters, we observe contours similar to the convex shape of $L_q$ priors with $q<1$ for the  peNMIG and NBPSS priors which implies weak shrinkage of large effects while small coefficients are strongly shrunken to zero.
 \item For the case of shared $\tau^2$, the shapes of the contours imply simultaneous shrinkage of both parameters instead of the strong shrinkage towards the coordinate axes observed for distinct importance parameters. This is exactly the desired type of shrinkage for parameters belonging to one effect $f(\nuvec$) to completely remove the effect from the model specification.
 \item As already noted in Section~\ref{subsec:NBPSS}, the specification of the prior in \citet{schfah11} differs from ours insofar as they consider the mixed model decomposition of effects. Additionally, \citet{schfah11} use a bimodal prior for the standardized regression effects with modes at $+1$ and $-1$. This effectively bounds the coefficients away from zero and thus encourages sampling from one mode of the posterior, while we instead explore the full posterior. Consequently,  the conditional posterior of $\tilde\betavec$ of NBPSS is a standard normal distribution $p_{\mbox{\scriptsize{NBPSS}}}(x)=\ND(x;0,1)$, while the one of peNMIG is a mixture of two normals with modes, $p_{\mbox{\scriptsize{peNMIG}}}(x)=0.5\ND(x;1,1)+0.5\ND(x;-1,1)$. Taking the ratio yields
$$\frac{p_{\mbox{\scriptsize{peNMIG}}}(x)}{p_{\mbox{\scriptsize{NBPSS}}}(x)}>1\quad\Leftrightarrow\quad|x|>\cosh^{-1}(\exp(0.5))\approx 1.08,$$
which explains the slightly heavier tails of peNMIG in Figures~\ref{fig:marginaldensities1d} and~\ref{fig:marginaldensities2d}.
\end{itemize}

We also study the implied constraint regions for the marginal prior of function evaluations $f(\nuvec)=\bvec_{\nu}'\betavec$, which can be derived in complete analogy by utilising that $\bvec_{\nu}'\betatildevec\sim\ND(0,\bvec_{\nu}'\mK^{-}\bvec_{\nu})$ with a generalised inverse $\mK^{-}$. In contrast, the marginal prior for function evaluations for the parameter expanded prior of \citet{schfah11} is not numerically accessible since it involves a complex mixture of $2^D$ components (where $D$ is the dimension of $\betavec$) due to the bimodal prior for the elements of $\betatildevec$.
Figure~\ref{fig:marginaldensitiesf} depicts marginal densities for the effect $f(\nuvec)$ evaluated at one (left panel) and two (right hand panel) randomly chosen covariate values of a sequence of $n=100$ equidistant values in $[-\pi,\pi]$. The resulting design matrix $\mB$ is based on cubic Bayesian P-splines with $D=\dim(\betavec)=22$. Hence, the bivariate plot corresponds to the situation of one shared importance parameter since we are interested in shrinkage of the effect evaluations for the same effect at different covariate values. Qualitatively, the behaviour from the marginal densities of the regression coefficient is translated to the function evaluations, i.e.~we observe a peak in zero and simultaneous shrinkage.
\subsubsection{Tail Behaviour and Behaviour in the Origin}
Visually, the marginal prior for $\betavec$ features a distinct peak as shown in the previous section. We now investigate more closely, whether this spike is finite or infinite by considering the behaviour of
 $\left.p_\betavec(\betavec)\right|_{\betavec=\nullvec}$. Using Equation~\eqref{eq:marg:prior} we obtain
\begin{eqnarray*}
 \left.p(\betavec)\right|_{\betavec=\nullvec} &=& 2p_\betatildevec(\nullvec)\left(\int_0^1\underbrace{p_\tau(\tau)}_{\ge p_\tau(1)}\frac{1}{\tau}d\tau + \underbrace{\int_1^\infty p_\tau(\tau)\frac{1}{\tau}d\tau}_{\ge0}\right)\\
 &\ge& 2p_\tau(1)p_\betatildevec(\nullvec)\int_0^1\frac{1}{\tau}d\tau
 = 2p_\tau(1)p_\betatildevec(\nullvec)\left[\log(\tau)\right]_0^1 = \infty,
\end{eqnarray*}
and therefore the marginal prior for $\betavec$ indeed has an infinite spike in zero. Note that we have shown that the multivariate parameter expanded prior has a spike in zero, while \citet{schfah11} have only shown the result for the univariate marginal prior. An infinite spike in zero is considered to induce particularly beneficial shrinkage properties since we obtain heavy penalisation of small effects.

The tail behaviour of the marginal prior for $\betavec$ can be studied by looking at the score function of $p(\betavec)$ which consists of the elements
\[
 \frac{\partial}{\partial\beta_d}p_\betavec(\betavec) = - \int p_\tau(\tau)p_\betatildevec(\betavec/\tau) \frac{\beta_d}{\tau^2}\frac{1}{|\tau|}d\tau.
\]
Figure~\ref{fig:scorefun} visualizes the resulting score function and compares it to the score function of the NMIG and peNMIG priors. From the graphical representation we find that all three prior structures have heavy tails such that the score functions are re-descending (i.e.~they approach zero as their argument tends to infinity) which induces Bayesian robustness of the resulting estimates. The score functions of the peNMIG and NBPSS priors resemble the shape of $L_q$ priors with $q$ close to zero, while the shape of the score function for the NMIG prior shows a more complex non-monotonously shape around zero.
\subsection{Propriety of the Posterior Distribution}\label{propriety}

While in Section~\ref{sec:BayVarSelDisReg} we do not explicitly change the design matrices to remove the nullspace of the precision matrices $\mK_{j,k}$ (both effects with NBPSS prior and the ones not under selection), we do derive an explicit mixed model representation of the predictors $\etavec_k$ in \eqref{eq:M2} in this section as this greatly simplifies the derivation of sufficient conditions for the propriety of the posterior.  As the exact conditions are also dependent on the prior structures employed, we  need to be more precise here about $\etavec^{\mbox{\scriptsize{in}}}$ and will therefore introduce a slightly different notation compared to that in Section~\ref{sec:BayVarSelDisReg}.
\subsubsection{Mixed Model Representation}
Assume we have $L_k$ effects in $\etavec_k^{\mbox{\scriptsize{in}}}$ and $J_k$ effects under selection and let furthermore $\etavec_k=\etavec_k^{\mbox{\scriptsize{in}}}+\etavec_k^{\mbox{\scriptsize{sel}}}$ be the complete predictors for $k=1,\ldots,K$ as defined in Section~\ref{subsec:pred}.

We then assume a mixed model type representation~\citep{FahKneLan2004} for $\etavec_k^{\mbox{\scriptsize{in}}}$
\[
\etavec_k^{\mbox{\scriptsize{in}}}=\sum_{l=1}^{L_k}\mZ_{l,k}^{\mbox{\scriptsize{in}}}(\tilde{\mU}_{l,k}^{\mbox{\scriptsize{in}}}\betavec_{\mathit{unpen},l,k}^{\mbox{\scriptsize{in}}}+\tilde{\mV}_{l,k}^{\mbox{\scriptsize{in}}}\betavec_{\mathit{pen},l,k}^{\mbox{\scriptsize{in}}})=\sum_{l=1}^{L_k}\left(\mU_{l,k}^{\mbox{\scriptsize{in}}}\betavec_{\mathit{unpen},l,k}^{\mbox{\scriptsize{in}}}+\mV_{l,k}^{\mbox{\scriptsize{in}}}\betavec_{\mathit{pen},l,k}^{\mbox{\scriptsize{in}}}\right)=\mU_{k}^{\mbox{\scriptsize{in}}}\betavec_{\mathit{unpen},k}^{\mbox{\scriptsize{in}}}+\mV_{k}^{\mbox{\scriptsize{in}}}\betavec_{\mathit{pen},k}^{\mbox{\scriptsize{in}}},
\]
where $\mU_{k}^{\mbox{\scriptsize{in}}}=(\mU_{1,k}^{\mbox{\scriptsize{in}}},\ldots,\mU_{L_k,k}^{\mbox{\scriptsize{in}}})$, $\mV_{k}^{\mbox{\scriptsize{in}}}=(\mV_{1,k}^{\mbox{\scriptsize{in}}},\ldots,\mV_{L_k,k}^{\mbox{\scriptsize{in}}})$, and $\betavec_{\mathit{unpen},k}^{\mbox{\scriptsize{in}}}=((\betavec_{\mathit{unpen},1,k}^{\mbox{\scriptsize{in}}})',\ldots,(\betavec_{\mathit{unpen},L_k,k}^{\mbox{\scriptsize{in}}})')'$, $\betavec_{\mathit{pen},k}^{\mbox{\scriptsize{in}}}=((\betavec_{\mathit{pen},1,k}^{\mbox{\scriptsize{in}}})',\ldots,(\betavec_{\mathit{pen},L_k,k}^{\mbox{\scriptsize{in}}})')'$. The columns of $\tilde{\mU}_{l,k}^{\mbox{\scriptsize{in}}}$ are a basis of $\ker(\mK_{l,k}^{\mbox{\scriptsize{in}}})$,  $\tilde\mV_{l,k}^{\mbox{\scriptsize{in}}}$ forms a basis of the images of $\mK_{l,k}^{\mbox{\scriptsize{in}}}$, such that $\dim(\betavec_{\mathit{pen},l,k}^{\mbox{\scriptsize{in}}})=\rank(\mK_{l,k}^{\mbox{\scriptsize{in}}})=\kappa_{l,k}^{\mbox{\scriptsize{in}}}$ and $\betavec_{\mathit{pen},l,k}^{\mbox{\scriptsize{in}}}|(\tau_{l,k}^2)^{\mbox{\scriptsize{in}}}\sim\ND(0,(\tau_{l,k}^2)^{\mbox{\scriptsize{in}}}\mI)$, while $\betavec_{\mathit{unpen},l,k}^{\mbox{\scriptsize{in}}}$ has dimension $D_{l,k}^{\mbox{\scriptsize{in}}}-\kappa_{l,k}^{\mbox{\scriptsize{in}}}$ and a flat prior. As a consequence, we obtain $L_k$ variance parameters $(\tau_{l,k}^2)^{\mbox{\scriptsize{in}}}$ for the $L_k$ penalized vectors of coefficients  $\betavec_l^{\mbox{\scriptsize{in}}}$ in $\etavec_k^{\mbox{\scriptsize{in}}}$.

For effects in $\etavec^{\mbox{\scriptsize{sel}}}$ we proceed similarly but with proper NBPSS priors on both parts of $\mK_{j,k}^{\mbox{\scriptsize{sel}}}$, $\rank(\mK_{j,k}^{\mbox{\scriptsize{sel}}})=\kappa_{j,k}^{\mbox{\scriptsize{sel}}}$ representing a basis of the nullspace and the image each. Hence, by construction all effects  under selection (after centring) can be assumed to have proper prior distributions. For non-linear effects of continuous covariates  with random walk priors of order $>2$ for instance, this is achieved by separating the polynomial parts up to order-1 and to include separate NBPSS prior on these, see Section~\ref{sec:BayVarSelDisReg} for details. We hence assume that the sub-predictors under selection are of the form
\begin{equation*}\begin{aligned}
\etavec_k^{\mbox{\scriptsize{sel}}}&=\sum_{j=1}^{J_k}\mV_{j,k}^{\mbox{\scriptsize{sel}}}\betavec_{j,k}^{\mbox{\scriptsize{sel}}}=\mV_{k}^{\mbox{\scriptsize{sel}}}\betavec_{k}^{\mbox{\scriptsize{sel}}},
\end{aligned}
\end{equation*}
where  $\mV_{k}^{\mbox{\scriptsize{sel}}}=(\mV_{1,k}^{\mbox{\scriptsize{sel}}},\ldots,\mV_{J_k,k}^{\mbox{\scriptsize{sel}}})$, and  $\betavec_{k}^{\mbox{\scriptsize{sel}}}=((\betavec_{1,k}^{\mbox{\scriptsize{sel}}})',\ldots,(\betavec_{J_k,k}^{\mbox{\scriptsize{sel}}})')'$. This yields $J_k$ importance parameters $(\tau_{j,k}^2)^{\mbox{\scriptsize{sel}}}$ with hyperparameters $\psi_{j,k}^2,\delta_{j,k},\omega_{j,k}$ in addition to the $J_k$ regression coefficients with NBPSS priors after re-parameterisation. We furthermore introduce $\kappa_k=\sum_{l=1}^{L_k}\kappa_{l,k}^{\mbox{\scriptsize{in}}}+\sum_{j=1}^{J_k}\kappa_{j,k}^{\mbox{\scriptsize{sel}}}$.

Finally, the complete predictor can  be written as
\begin{equation}\label{eq:2}
\etavec_k=\mU_k^{\mbox{\scriptsize{in}}}\betavec_{\mathit{unpen},k}^{\mbox{\scriptsize{in}}}+\mV_k^{\mbox{\scriptsize{in}}}\betavec_{\mathit{pen},k}^{\mbox{\scriptsize{in}}}+\mV_{k}^{\mbox{\scriptsize{sel}}}\betavec_{k}^{\mbox{\scriptsize{sel}}}=\mU_k\betavec_{\mathit{unpen},k}+\mV_k\betavec_{\mathit{pen},k},
\end{equation}
where we denote $\betavec_{\mathit{unpen},k}\equiv\betavec_{\mathit{unpen},k}^{\mbox{\scriptsize{in}}}$, $\betavec_{\mathit{pen},k}=((\betavec_{\mathit{pen},k}^{\mbox{\scriptsize{in}}})',(\betavec_{k}^{\mbox{\scriptsize{sel}}})')'$, $\mU_k\equiv\mU_k^{\mbox{\scriptsize{in}}}$, $\mV_k=(\mV_{k}^{\mbox{\scriptsize{in}}},\mV_{k}^{\mbox{\scriptsize{sel}}})$.

Let us in the sequel assume that the matrices $\mU_k$ have full column rank $r_k$, $k=1,\ldots,K$ and define for $\mX_k=(\mU_k,\mV_k)$ and
$t_k=\rank(\mX_k)-\rank(\mU_k)\le\dim(\betavec_{\mathit{pen},k})$,
\begin{equation}\label{eq:meanreg:ranks}
\rank(\mX_k)=r_k+t_k. 
\end{equation}

\begin{remark}\label{remark:ranks}
\
In order to obtain a full column rank matrix of unpenalised effects  in the mixed model representation~\eqref{eq:2}, all superfluous columns have to be deleted. In particular, duplicated constant columns representing the levels of the functions are deleted which is a simple way to include the centring restrictions and is equivalent to the centring of functions that we include in our MCMC algorithm. Furthermore, using the one-to-one relationship between original parameterisation and the reparameterised model the restrictions for one presentation can be deduced from the other one. Hence, sufficient rank conditions can be formulated directly for the reparameterised model (\ref{eq:2}) and do not have to be traced back to the original parameterisation, see~\citet[][Remark 1]{KleKne2016} for a detailed derivation of this result.
\end{remark}

\subsubsection{Conditional Independence Assumptions}
To derive the posterior distribution of model~\eqref{eq:M1} to~\eqref{eq:M5}, we make the usual conditional independence assumptions (see the Online Appendix~A.1, conditions {(a.1)--(a.3b)}) by labelling for $k=1,\ldots K$ the coefficients $\betavec_{l,k}^{\mbox{\scriptsize{in}}}$ with variances $(\tau_{l,k}^2)^{\mbox{\scriptsize{in}}}$, $l=1,\ldots,L_k$  for effects not under selection; and $\betavec_{j,k}^{\mbox{\scriptsize{sel}}}$, $(\tau_{j,k}^2)^{\mbox{\scriptsize{sel}}}$, $\psi_{j,k}^2$, $\delta_{j,k}$, $\omega_{j,k}$, ,$J=1,\ldots,J_k$, for the  effects with NBPSS prior. In general, they mean that priors for different effects are assumed to be independent, while within an effect they are dependent by construction.
 In general, prior independence assumptions
should be a reasonable working assumption which also does not rule out posterior
dependence. 
Note that we always assume proper NBPSS priors and in particular $a_{j,k}>0$, $b_{j,k}>0$ in the priors for $\psi_{j,k}^2$. This is justified by our considerations on prior elicitation as discussed in Section~3.2 of the main paper.
In the following we assume that conditions (a.1)--(a.3b) of the Online Appendix~A.1 hold.

\subsubsection{Gaussian Mean Regression}\label{sec:gaussian-mean-regression}
Assume in this section a Gaussian mean regression model for $\yvec=(y_1,\ldots,y_n)'$ with predictor $\etavec$ from~\eqref{eq:2} in mixed model representation, i.e.
\begin{equation}\label{eq:meanreg2}
\yvec=\etavec+\varepsilonvec,\quad \varepsilonvec\sim\ND(\nullvec,\tau_{\varepsilon}^2\mI_{n}),
\end{equation}
where we assume
\[
p(\tau_{\varepsilon}^2)\propto \frac{1}{(\tau_{\varepsilon}^2)^{a_{\varepsilon}+1}}\exp\left(-\frac{b_{\varepsilon}}{\tau_{\varepsilon}^2}\right)
\]
for the error variance. Note that $k=1$ in this subsection and that $J_k,L_k,\kappa_{k}$ are replaced by $J,L,\kappa$. Applying the mixed model representation~\eqref{eq:2} allows us writing~\eqref{eq:meanreg2} as
\[
\yvec=\mU\betavec_{\mathit{unpen}}+\mV\betavec_{\mathit{pen}}+\varepsilonvec,
\]
and with the corresponding rank assumptions from above.

\emph{b.~Conditions for Gaussian Mean Regression} 
\begin{Lalign}
\tag{b.1}\label{eq:B1} & a_l^{{\mbox{\scriptsize{in}}}}<b_l^{\mbox{\scriptsize{in}}}=0 \mbox{ or } b_l^{\mbox{\scriptsize{in}}}>0,\; l=1,\ldots,L.\\
\tag{b.2}\label{eq:B2} &\kappa_l^{{\mbox{\scriptsize{in}}}}+2 a_l^{{\mbox{\scriptsize{in}}}}>0,\,l=1,\ldots,L.&\\
\tag{b.3}\label{eq:B3} &\kappa_l^{{\mbox{\scriptsize{in}}}}+2 a_l^{{\mbox{\scriptsize{in}}}}>\kappa-t,\,l=1,\ldots,L.&\\
\tag{b.4}\label{eq:B4} &\kappa_j^{{\mbox{\scriptsize{sel}}}}+2a_j^{{\mbox{\scriptsize{sel}}}}-1>\kappa-t,\,j=1,\ldots,J.&\\
\tag{b.5}\label{eq:B5} &n+2a_{\varepsilon}+2\sum_{l=1}^L a_l^{{\mbox{\scriptsize{in}}}}>r+J.&
\end{Lalign}
\begin{Lalign}
\tag{b.6}\label{eq:B6} &n+2a_{\varepsilon} +2\sum_{l=1}^L \min(0,a_l^{{\mbox{\scriptsize{in}}}})>r+J.&\\
\tag{b.7}\label{eq:B7} &\text{SSE}+2 b_{\varepsilon}>0.&
\end{Lalign}

Condition~\eqref{eq:B1} excludes Jeffrey's prior (corresponding to $a_l^{{\mbox{\scriptsize{in}}}}=b_l^{{\mbox{\scriptsize{in}}}}=0$) for effects not under selection but allows for flat priors on variances and standard deviations $(\tau_l^2)^{\mbox{\scriptsize{in}}}$. Conditions~\eqref{eq:B2} to~\eqref{eq:B4} relate the ranks $\kappa_l^{\mbox{\scriptsize{in}}}$ and $\kappa_j^{\mbox{\scriptsize{sel}}}$ of the prior precision matrices of each of the effects to the rank $\kappa$ of all prior precision matrices. For effects not under selection, the conditions can be ensured by increasing  $a_l^{{\mbox{\scriptsize{in}}}}$. Condition~\eqref{eq:B5} restricts the number of all effects to be smaller or equal to the number of observations but can be relaxed by increasing the hyperparameters values $a_{\varepsilon}$ and $a_l^{\mbox{\scriptsize{in}}}$. Condition~\eqref{eq:B7} is always fulfilled for $b_{\varepsilon}>0$. In case of an improper prior for $\tau_{\varepsilon}^2$, $\SSE>0$ has to be assured, while $b_{\varepsilon}>0$ becomes necessary when the number of unknown parameters is greater than $n$.

\begin{theo}
\label{theo:propriety:gaussian}
Consider the Gaussian mean regression model~\eqref{eq:meanreg2} with mixed model representation~\eqref{eq:2} and rank conditions from~\eqref{eq:meanreg:ranks}.
\begin{enumerate}
\item $\kappa=t$: Then, conditions~\eqref{eq:B1},\eqref{eq:B3},\eqref{eq:B5} and \eqref{eq:B7} are necessary for the propriety of the joint posterior while conditions~\eqref{eq:B1},\eqref{eq:B3},\eqref{eq:B4},\eqref{eq:B6} and~\eqref{eq:B7} are sufficient for the propriety of the joint posterior.
\item $\kappa<t$: Then, conditions~\eqref{eq:B1},\eqref{eq:B2},\eqref{eq:B5} and \eqref{eq:B7} are necessary for the propriety of the joint posterior while conditions~\eqref{eq:B1},\eqref{eq:B3},\eqref{eq:B4},\eqref{eq:B6} and~\eqref{eq:B7} are sufficient for the propriety of the joint posterior.
\end{enumerate}
\end{theo}
The proof of Theorem~\ref{theo:propriety:gaussian} is given in  the Online Appendix~A.3.

\begin{remark}\label{remark:meanreg:comparisonIG}
\
For effects not under selection, additional conditions on the ranks $\kappa_l$ and the number of effects compared to the shape parameters $(a_l)^{{\mbox{\scriptsize{in}}}}$ of the priors are required, as the latter can be improper and hence  $(a_l)^{{\mbox{\scriptsize{in}}}}<0$ becomes possible. Consequently, one has to consider the cases $t=\kappa$ or $L=1$ as well as $t<\kappa$ and  $L>1$ separately. This is not necessary for effects with NBPSS prior.
\end{remark}

\subsubsection{Distributional Regression}\label{sec:distributional-regression}

In order to achieve sufficient conditions for the propriety of the posterior in distributional regression, we define a normalized submodel with Gaussian errors to be able to apply results of Theorem~1. More precisely, we first separate the random effect with largest dimension in each predictor of~\eqref{eq:2}, such that we obtain
\[
 \etavec_{k} = \mU_{k}^{\mbox{\scriptsize{in}}}\betavec_{\mathit{unpen},k}^{\mbox{\scriptsize{in}}} + \mV_{k}\bvec_{k} + \mV_{\varepsilon,k}\bvec_{\varepsilon,k},
\]
where $\mV_{\varepsilon,k}\bvec_{\varepsilon,k}$ corresponds to the effect with proper prior and with the largest dimension, $\dim(\bvec_{\varepsilon,k})=\rank(K_{\varepsilon,k})=\kappa_{\varepsilon,k}$, and $\mV_{k}\bvec_{k}$ contains all remaining  effects with proper prior, both the ones with NBPSS prior and the ones not under selection with usual inverse gamma priors. Note that $\bvec_{k}$ is based on $J_k^{\ast}=L_k+J_k-1$ effects in the notation in~\eqref{eq:2}, with $\kappa_k$ denoting the sum of ranks of the $J_k^{\ast}$ precision matrices of predictor $k$, and where, w.l.o.g.~we assume that the effects in the predictors are ordered such that the $(J_k+L_k)$-th effect corresponds to the random effect in the mixed model representation with largest dimension. Similarly, the design matrix $(\mV_{k},\mV_{\varepsilon,k})$ corresponds to the design matrix $(\mV_{k}^{\mbox{\scriptsize{in}}},\mV_{k}^{\mbox{\scriptsize{sel}}})$. Note also, that $\bvec_{\varepsilon,k}$ can originate from an effect not under selection or one with NBPSS prior and we distinguish the two cases in Theorems~\ref{theo:propriety:distributional} and~\ref{theo:propriety:distributional2}.

Assume that the set of observations can (after re-ordering) be partitioned such that for $n^\ast\ge1$
\begin{Lalign}
\tag{c.1}\label{eq:C1} &\displaystyle\int\ldots\int p(y_i|\eta_{i1},\ldots,\eta_{iK})\mathrm{d}\eta_{i1}\ldots\mathrm{d}\eta_{iK}<\infty \mbox{ for } i=1,\ldots,n^\ast.&\\
\tag{c.2}\label{eq:C2}& p(y_i|\eta_{i1},\ldots,\eta_{iK})\leq M\mbox{ for }i=n^\ast+1,\ldots,n,&
\end{Lalign}
where $\eta_{ik}=h_k^{-1}(\vartheta_{ik})$, $i=1,\ldots,n$, $k=1,\ldots,K$.
This implies that for at least one observation the density is integrable (with respect to the predictors) and that all remaining densities are bounded. For discrete distributions, all densities are automatically bounded by 1 so that only Condition~\eqref{eq:C1} can be an issue in practice. Condition~\eqref{eq:C1} is usually fulfilled if certain restrictions apply on specific parameters that exclude extreme values on the boundary of the parameter space, see~\citet{KleKneLan2015} for a more detailed discussion on count data and binary distributions. For continuous distributions, the densities are sometimes not bounded (e.g.~for the gamma distribution). Note that this is not a problem when all observations fulfil Condition~\eqref{eq:C1} since $n^{\ast}=n$ is allowed. Similar as for the discrete distributions, integrability of the densities can be assured by the assumption that none of the distributional parameters is on the boundary of the parameter space (an assumption that would also have to be made to apply standard maximum likelihood asymptotics).

Let $\tilde{n}_{\varepsilon}=\min\{\kappa_{\varepsilon,1},\ldots,\kappa_{\varepsilon,K}\}$ and assume that we can choose $\tilde{n}_{\varepsilon}$ observations including at least one observation fulfilling~\eqref{eq:C1} to define the submodel
\begin{equation}\label{eq:distributiona:mixedmodel}
 \etavec_{k,s} = \mU_{k,s}\betavec_{\mathit{unpen},k} + \mV_{k,s}\bvec_{k} + \mV_{\varepsilon,k,s}\bvec_{\varepsilon,k}
\end{equation}
with these observations, such that $\mV_{\varepsilon,k,s}\bvec_{\varepsilon,k}\sim\ND(\nullvec, \tau_{\varepsilon,k}^2\mV_{\varepsilon,k,s}\mV_{\varepsilon,k,s}{}')$. Then the following rank conditions have to be fulfilled:
\begin{Lalign}
\tag{c.3}\label{eq:C3} &\mbox{The design matrix }\mU_{k,s}\mbox{ has full rank }r_{k}.&\\
\tag{c.4}\label{eq:C4}&\rank(\mU_{k},\mV_{k})=\rank(\mU_{k,s},\mV_{k,s})=r_{k}+t_{k}.&\\
\tag{c.5}\label{eq:C5}&\rank(\mV_{\varepsilon,k,s})=\tilde{n}_{\varepsilon}\mbox{ i.e.~}\mV_{\varepsilon,k,s}\mbox{ is of full rank for } k=1,\ldots,K.&
\end{Lalign}
To ensure~\eqref{eq:C3}, superfluous columns arising from the reparameterisation have to be deleted. In particular, duplicated constant columns representing the levels of the functions are deleted, see~\citet[][Remark~2~(iii) for details]{KleKne2016}. Condition~\eqref{eq:C4} indicates that the rank of the design matrices in the submodel is the same as in the complete model whereas~\eqref{eq:C5} defines a similar restriction for the design matrix of the largest random effect arising from the mixed model representation.
Finally, the  normalised submodel
\begin{equation}\label{eq:submodnormal}
 \etatildevec_{k,s} = \tilde{\mU}_{k,s}\betavec_{\mathit{unpen},k} + \tilde\mV_{k,s}\bvec_{k} + \varepsilonvec_{k,s},\quad\varepsilonvec_{k,s}\sim\ND(\nullvec,\tau_{\varepsilon,k}^2\mI_{\tilde{n}_{\varepsilon}})
\end{equation}
 is obtained by multiplying~\eqref{eq:distributiona:mixedmodel} with $\mM_{k}=(\mV_{\varepsilon,k,s}\mV_{\varepsilon,k,s}{}')^{-1/2}$ such that
 $\etatildevec_{k,s}=\mM_k\etavec_{k,s},$ $\mtildeU_{k,s}=\mM_k\mU_{k,s}$, $\mtildeV_{k,s}=\mM_k\mV_{k,s}$, and $\varepsilonvec_{k,s}$ represents an i.i.d.~random effect.

The corresponding residual sum of squares for the normalised submodel is
\begin{eqnarray}\label{sse}
\SSE_{k,s}:= \left(\etatildevec_{k,s}-\tilde\mU_{k,s}\betavec_{\mathit{unpen},k}-\tilde\mV_{k,s}\bvec_k\right)'\left(\etatildevec_{k,s}-\tilde\mU_{k,s}\betavec_{\mathit{unpen},k}-\tilde\mV_{k,s}\bvec_k\right).
\end{eqnarray}

To derive sufficient conditions for the propriety of the posterior we have to distinguish two cases: the largest random effect $\varepsilon_{k,s}$ corresponds to an effect with a) NBPSS prior and b) not under selection and with the usual inverse gamma priors for the variance $\tau_{\varepsilon,k}$.
\vspace{-1.0mm}
\begin{Lalign}
\tag{c.6a}\label{eq:C6a} & a_{l,k}^{{\mbox{\scriptsize{in}}}}<b_{l,k}^{\mbox{\scriptsize{in}}}=0 \mbox{ or } b_{l,k}^{\mbox{\scriptsize{in}}}>0,\; l=1,\ldots,L_k.\\
\tag{c.6b}\label{eq:C6b} & a_{l,k}^{{\mbox{\scriptsize{in}}}}<b_{l,k}^{\mbox{\scriptsize{in}}}=0 \mbox{ or } b_{l,k}^{\mbox{\scriptsize{in}}}>0,\; l=1,\ldots,L_k-1.\\
\tag{c.7a}\label{eq:C7a} &\kappa_{l,k}^{{\mbox{\scriptsize{in}}}}+2 a_{l,k}^{{\mbox{\scriptsize{in}}}}>\kappa_{k}-t_k,\,l=1,\ldots,L_k.&\\
\tag{c.7b}\label{eq:C7b} &\kappa_{l,k}^{{\mbox{\scriptsize{in}}}}+2 a_{l,k}^{{\mbox{\scriptsize{in}}}}>\kappa_{k}-t_k,\,l=1,\ldots,L_k-1.&\\
\tag{c.8a}\label{eq:C8a} &\kappa_{j,k}^{{\mbox{\scriptsize{sel}}}}+2a_{j,k}-1>\kappa_{k}-t_k,\,j=1,\ldots,J_k-1.&\\
\tag{c.8b}\label{eq:C8b} &\kappa_{j,k}^{{\mbox{\scriptsize{sel}}}}+2a_{j,k}-1>\kappa_{k}-t_k,\,j=1,\ldots,J_k.&\\
\tag{c.9a}\label{eq:C9a} &\tilde{n}_{\varepsilon}+2a_{\varepsilon,k} +2\sum_{l=1}^{L_k} \min(0,a_{l,k}^{{\mbox{\scriptsize{in}}}})>r_k+(J_k-1).&\\
\tag{c.9b}\label{eq:C9b} &\tilde{n}_{\varepsilon}+2a_{\varepsilon,k} +2\sum_{l=1}^{L_k-1} \min(0,a_{l,k}^{{\mbox{\scriptsize{in}}}})>r_k+J_k.&\\
\tag{c.10a}\label{eq:C10a} &\text{ SSE}_k>0.&\\
\tag{c.10b}\label{eq:C10b} &\text{ SSE}_k+2 b_{\varepsilon}>0.&
\end{Lalign}

Above, Conditions (c.$\cdot$a) each correspond to the case that the largest random effect has a variance with inverse gamma prior, while  Conditions (c.$\cdot$b) each are active when the variance of the largest random effect has an NBPSS prior. Conditions~\eqref{eq:C6a},\eqref{eq:C6b} require that if for effects not under selection $b_{l,k}^{\mbox{\scriptsize{in}}}$
 is set to zero, the parameter
$a_{l,k}^{\mbox{\scriptsize{in}}}$ has to be negative. This includes situations corresponding to
flat priors for the random effects variance ($a_{l,k}^{\mbox{\scriptsize{in}}}=-1$)
or standard deviation ($a_{l,k}^{\mbox{\scriptsize{in}}}=-0.5$) but excludes Jeffreys’
prior ($a_{l,k}^{\mbox{\scriptsize{in}}}$=0).
Conditions~\eqref{eq:C7a}, \eqref{eq:C7b} and~\eqref{eq:C8a}, \eqref{eq:C8b} relate the rank of the random effects part of
one individual effect to the sum of all rank deficiencies in the
corresponding predictor, are similar for effects not under selection and the ones with NBPSS prior and require that the dimensionality is
not too small. The condition can be ensured by increasing the
shape parameters $a_{l,k}^{{\mbox{\scriptsize{in}}}}$ and $a_{j,k}$, respectively.
Conditions~\eqref{eq:C9a}, \eqref{eq:C9b} restrict the number
of effects not under selection and with flat prior to be at most equal to the dimension
of the largest random effects part in the model but can again be relaxed by increasing the shape parameters $a_{l,k}^{{\mbox{\scriptsize{in}}}}$.
Finally, Conditions~\eqref{eq:C10a}, \eqref{eq:C10b} require
that there is variation in the residual sum of squares in the
normalized submodel (implying that not all effects are zero) in situations where the largest random effect has an NBPSS prior and either variation in the residual sum of squares or
$b_{\varepsilon}>0$ when the largest random effect has the usual inverse gamma prior on the variances. The latter requirement can always be ensured in
practice but excludes flat priors for the random effects variances
or standard deviations.

\begin{theo}
\label{theo:propriety:distributional}
Consider the distributional regression model with densities~\eqref{eq:M1} and predictors~\eqref{eq:M2}. Let $\varepsilon_{k,s}$ be an i.i.d.~random effect with variance $\tau_{\varepsilon,k}^2\sim\IGD(a_{\varepsilon,k},b_{\varepsilon,k})$. Then, Conditions~\eqref{eq:C1},~\eqref{eq:C2} on the densities,~\eqref{eq:C3} to~\eqref{eq:C5} on the ranks as well as~\eqref{eq:C6b}, \eqref{eq:C7b}, \eqref{eq:C8b}, \eqref{eq:C9b}, \eqref{eq:C10b} on the hyperpriors  are sufficient conditions for a proper posterior.
\end{theo}
The proof of Theorem~\ref{theo:propriety:distributional} follows from the proof of~\cite{KleKneLan2015} using Theorem~\ref{theo:propriety:gaussian} above as we assume that all NBPSS priors are proper.

\begin{theo}
\label{theo:propriety:distributional2}
Consider the distributional regression model with densities~\eqref{eq:M1} and predictors~\eqref{eq:M2}. Let $\varepsilon_{k,s}$P be an i.i.d.~random effect with NBPSS prior with parameters $\tau_{\varepsilon,k}^2\sim\GaD(1/2,1/(2r(\delta_{\varepsilon,k})\psi_{\varepsilon,k}^2))$, $\psi_{\varepsilon,k}^2\sim\IGD(a_{\varepsilon,k},b_{\varepsilon,k})$, $\delta_{\varepsilon,k}\sim\BeD(\omega_{\varepsilon,k})$, $\omega\sim\BetaD(a_{0,\varepsilon,k},b_{0,\varepsilon,k})$. Then, Conditions~\eqref{eq:C1},~\eqref{eq:C2} on the densities,~\eqref{eq:C3} to~\eqref{eq:C5} on the ranks as well as~\eqref{eq:C6a}, \eqref{eq:C7a}, \eqref{eq:C8a}, \eqref{eq:C9a}, \eqref{eq:C10a} on the hyperpriors  are sufficient conditions for a proper posterior.
\end{theo}
The proof of Theorem~\ref{theo:propriety:distributional2} is given in the  Online Appendix~A.4.

\section{Posterior Estimation}\label{sec:Inf}\label{subsec:implement}
\paragraph{Update of the Basis Coefficients.} Due to the modular structure of Markov chain Monte Carlo (MCMC) simulation algorithms, no changes in the MCMC scheme developed by \citet{KleKneLanSoh2015} are required for updating the basis coefficients $\betavec$ when supplementing them with a NBPSS prior instead of the standard inverse gamma prior. We therefore apply iteratively weighted least squares based approximations to the log full conditional and generate proposals from the multivariate normal distribution $\ND(\muvec, \mP^{-1})$ with expectation and precision matrix given by
\begin{equation}\label{eq:prop}
 \muvec = \mP^{-1}\mB'\mW(\ytildevec-\etavec_{-}) \qquad \mP = \mB'\mW\mB+\frac{1}{\tau^2}\mK
\end{equation}
where $\etavec_{-}=\etavec-\mB\betavec$ is the predictor without the effect currently updated and the working observations $\ytildevec$ and weights $\mW$ are determined based on first and second derivatives of the log-likelihood with respect to the predictor.
\paragraph{Update of the Smoothing Variance for Effects not Subject to Selection.} For effects not subject to selection, we consider an inverse gamma prior $\tau^2\sim\IGD(a,b)$ for the smoothing variances such that the update of $\tau^2$ can be done via a simple Gibbs sampling step drawing from $\tau^2|\cdot\sim\IGD(a',b')$, with updated parameters $a'=\frac{\rank(\mK)}{2}+a$, $b'=\frac{1}{2}\betavec'\mK\betavec+b$.
\paragraph{Update of the Squared Importance Parameter for Effects Subject to Selection.} The full conditional $p(\tau^2|\betavec,\delta,\psi^2)$ is a generalised inverse Gaussian distribution $\mbox{GIG}(p,q,c)$, with $p=-0.5\rank(\mK)+0.5$, $q=1/(r(\delta)\psi^2)$, $c=\betavec'\mK\betavec$ and can be generated efficiently in a Gibbs-step. This has the advantage that $\tau^2$ can be generated independently of the likelihood in an efficient Gibbs step. This is no longer possible when the prior is formulated for the importance parameter $\tau$ as in~\citep{schfah11} where a Metropolis-Hastings update is required, see  the Online Appendix~C.
\paragraph{Updates for the Hyperparameters of the NBPSS prior.} For the hyperparameters of the NBPSS prior, we obtain Gibbs sampling steps via the following full conditionals:
\begin{itemize}
\item Inclusion indicator $\delta$:
\[
 p(\delta=1|\cdot) = \frac{1}{1+\frac{\varphi(\tau;0; r\psi^2) (1-\omega)}{\varphi(\tau;0;\psi^2)\omega} }=\frac{1}{1+\frac{1-\omega}{\omega} L},
\]
where $\varphi(\cdot; \mu, \sigma^2)$ denotes the density of the normal distribution with mean $\mu$ and variance $\sigma^2$ and
\[
 L=\frac{ \varphi(\tau; 0, r\psi^2)}{\varphi(\tau;0, \psi^2)} =\frac{1}{ \sqrt{r}}e^{-\frac{\tau^2}{2\psi^2}(1/r-1)}.
\]
\item Hyper-variance $\psi^2$:
\[
 \psi^2| \cdot \sim \IGD\left(a+0.5,b+\frac{\tau^2}{2r(\delta)}\right)
\]
\item Inclusion probability $\omega$:
\[
 \omega|\cdot \sim \BetaD(a_0+\delta,b_0+1-\delta)
\]
\end{itemize}
Note that it is also possible to use the same $\omega$ for multiple effects simultaneously. If $\omega$ relates to a total of $L$ effects, the full conditional is then given by
\[
 \omega|\cdot \sim \BetaD\left(a_0+\sum_{l=1}^L\delta_l,b_0+L-\sum_{l=1}^L\delta_l\right)
\]
\paragraph{Implementation.} Spike and slab based effect selection in distributional regression has been implemented in a developer version of BayesX~\citep{BelBreKleKneLanUml2015} which is available from the authors on request. The software makes use of methods for efficient storing of large data sets and sparse matrix algorithms for sampling from multivariate Gaussian distributions~\citep{GeoLiu1981,Rue2001} and also allows us to access existing procedures for example for computing simultaneous confidence bands for nonparametric effects as developed in~\citet{KriKneCla2010}. Hyperparameter elicitation is integrated in the R-package \texttt{sdPrior}~\citep{sdPrior}.

\section{Empirical Evaluations}
\subsection{Simulations}\label{sec:sims}
To evaluate the performance of the NBPSS prior for effect selection in distributional regression, we conducted extensive simulations under various settings. We distinguish different scenarios for the predictor complexity, models including and excluding spatial effects, four selected response distributions, varying sample sizes, correlated and uncorrelated covariates and a set of user-defined parameters for hyperprior elicitation.  Specifically,
 \begin{itemize}
 \item we consider Gaussian responses with effects only on the expectation, a Gaussian location-scale model, Poisson regression and zero-inflated Poisson models.
 \item we specify four test functions\vspace{-0.6cm}
 \begin{multicols}{2}
 \begin{itemize}
\item $f_1(x) = x$
 \item $f_2(x) = x + \frac{(2x-2)^2}{5.5}$
  \item $f_3(x) = -x + \pi \mathrm{sin}(\pi x)$
\item $f_4(x) = 0.5x + 15\phi(2(x-0.2))-\phi(x+0.4)$.
\end{itemize}
\end{multicols}\vspace{-0.25cm}
\item we distinguish two scenarios in terms of the predictor complexity:
\begin{itemize}
\item \textbf{low sparsity} in which out of 16 included covariates 12 have non-zero influence. The true linear predictor is $\eta = f_1(x_1)+f_2(x_2)+f_3(x_3) +f_4(x_4)
 	   +1.5\left(f_1(x_5)+f_2(x_6)+f_3(x_7)+f_4(x_8)\right)
 	+2(f_1(x_9) +f_2(x_{10})+f_3(x_{11})+f_4(x_{12})$
and we simulate the two cases with additional and without additional spatial effect $f_{spat}(s)$, labeled as \lq spatial/non-spatial\rq. These settings are used for $\eta_{\mu}$  in the homoscedastic Gaussian and the  Gaussian location-scale model, as well as  for $\eta{_\lambda}$ in the Poisson and the zero-inflated Poisson model.
\item \textbf{high sparsity} in which out of eight included covariates four have non-zero influence. The true linear predictor is $\eta = f_1(x_1)+f_2(x_2)+f_3(x_3) +f_4(x_4)$ and we again simulate the two cases with additional and without additional spatial effect $f_{spat}(s)$. These settings are used for $\eta_{\sigma^2}$  in the Gaussian  location-scale model and  for $\eta{_\pi}$ in the zero-inflated Poisson model.
\end{itemize}
\item we generate covariates either
\begin{itemize}
\item as i.i.d.~realizations from $U[-2,2]$ or
\item from an $AR(1)$ process with correlation $\rho =0.7$
\end{itemize}
and standarize $x$ in order to facilitate prior elicitation.
\item we simulate 150 replications for each combination of the settings.
\item we use six combinations of $\alpha$ and $c$ for the elicitation of the prior hyperparameters $b$ and $r$ arising from the pairwise combination of
\begin{itemize}
\item $\alpha = 0.05, 0.1, 0.2,$
\item $c = 0.1, 0.2$.
\end{itemize}
\item we consider the sample sizes $n=200;1,000$ for Gaussian, $n=500;2,000$ for Poisson, $n=1,000;2,000$ for Gaussian location-scale and zero-inflated Poisson responses. The sample sizes have been chosen to reflect a challenging (small sample size) and a relatively informative (large sample size) setting, taking the different complexity of the model structures into account.
\end{itemize}
As a competitor for the single parameter distributions Gaussian and Poisson, we consider the peNMIG prior of~\citet{schfah11} implemented in the R-package \texttt{spikeSlabGAM}~\citep{spikeslabgam}. We refrain from comparison with further variable selection priors mentioned in the introduction as these usually lack applicability beyond the framework of generalized linear models. Hyperparameter elicitation for the NBPSS prior was performed with the package \texttt{sdPrior}~\citep{sdPrior} and estimation was done with the current developer version of BayesX \citep{BelBreKleKneLanUml2015}. For both the NBPSS and the peNMIG prior, non-linear effects are based on 20 cubic B-spline basis functions constructed from an equidistant set of knots combined with second-order random walk prior unless stated otherwise. 

In the following, we restrict ourselves to the main conclusions, a detailed description about simulation settings and evaluation including complete graphical evidence is provided in the Online Appendix~D. As a general outcome, the NBPSS prior results in very good performance for the selection of relevant effects even in challenging distributional regression settings with effect selection on multiple distributional parameters, where no competing Bayesian variable selection approach is available so far. Evidence for that is given in  Figures~\ref{fig:zip_lambda_pi} and~\ref{fig:logscores} showing posterior inclusion probabilities and the ratio between predictive NBPSS log-scores and oracle log-scores (i.e. log-scores arising from a model with given, true predictor specification), respectively, in the zero-inflated Poisson model. The log-scores have been computed from independently generated test data sets with 5,000 observations.

In the simple exponential family framework with only one single regression predictor, the NBPSS prior turns out to be a strong competitor to the peNMIG prior (see Figure~\ref{fig:acc_poisson} for overall accuracy results of the Poisson model). Selection of
large coefficient blocks such as spatial effects works well for all types of response distributions, while these are particularly problematic with~peNMIG due to severe mixing problems. On the other hand, the explicit reparameterisation of non-linear effects used with the peNMIG prior (as compared to the constrained sampling approach that NBPSS is based on) seems to have some advantages in separating the linear and non-linear part of non-linear effects in cases where the true effect is close to linear and at the same time covariates are strongly correlated.

Coinciding with previous evidence on Bayesian effect selection, we find a strong impact of hyperprior parameter choice on the resulting effect selection performance. Our interpretable yet flexible way of eliciting hyperprior parameters equips data analysts with an intuitive approach for choosing these hyperparameters. More precisely, changing the probability $\alpha$ and the threshold $c$ can help to balance between the true positive and false negative rates of effect selection. Choosing $\alpha$ and $c$ smaller, results in more conservative, i.e.~sparser models. Based on our simulations, we suggest $\alpha=c=0.1$ as default values in our applications.

In summary, our simulations demonstrate that the NPBSS prior provides a promising approach for Bayesian effect selection that extends existing methods to a framework that is applicable in any  distributional regression model comprising both multiple hierarchical predictor specifications and high-dimensional coefficient vectors. In addition, our effect decomposition allows to select the linear part and its non-linear deviation for an effect of a continuous covariate separately.

\subsection{Applications}\label{sec:App}
In this section, we demonstrate the efficacy of a simultaneous selection approach via the NBPSS prior specification and its applicability for non-Gaussian, discrete or multivariate data.
Core information about the different data sets \emph{Patents}, \emph{Nigeria} and \emph{House prices} including the type of response distribution, number of observations and effects can be found in Table~\ref{tab:summaries}.  Estimates shown in the subsequent subsections are all the model-averaged  estimates obtained from the MCMC iterates with the NBPSS prior and the covariates have been standardized for prior elicitation reasons.


\subsubsection{Number of Patent Citations}
The \emph{Patents} data set contains the number of citations
of patents granted by the European Patent Office (EPO). An
inventor who applies for a patent has to cite all related, already
existing patents his patent is based on.~\cite{KleKneLan2015} use this data set to illustrate their developed methodology on Bayesian zero-inflated and overdispersed count data and conducted variable selection in a stepwise forward approach based on the deviance information criterion (DIC).
In the following, we focus on zero-inflated Poisson (ZIP) models for analysing the number of patent citations. The ZIP model has two distributional parameters, $\lambda$, the rate of the count process, and $\pi$ the probability of observing an excess of zeros. Including all available variables in one of the predictors $\eta_k$, $k=1,2$ reads as
\[
\eta_k =\beta_{0,k}+ \xvec'\betavec+f_{1,k}(\mathit{year})+f_{2,k}(\mathit{ncountry}) + f_{3,k}(\mathit{nclaims}),
\]
where $\xvec$ contains the continuous variables $\mathit{year}$ (year when  patent was granted), $\mathit{ncountry}$ (number of designated states for  patent), $\mathit{nclaims}$ (number of patent claims), as well as the binary indicators $\mathit{ustwin}$ (twin patent in the US), $\mathit{opp}$ (oppositions against the patent), $\mathit{biopharm}$ (patent from the biotech/pharma sector), $\mathit{patus}$ (patent holder from the US) and $\mathit{patgsgr}$ (patent holder from Germany, Switzerland or Great Britain), see~Table E.1 in the Online Appendix for summary statistics of the variables. Possible non-linear effects of the three continuous variables are captured by the functions $f_1$ to $f_3$. The predictor specifications of the model identified in \citet{KleKneLan2015} via stepwise DIC-selection are
\begin{equation*}\begin{aligned}
\eta_{\lambda} &=\beta_{0,\lambda}+ \beta_{1,\lambda}\mathit{opp}+\beta_{2,\lambda}\mathit{biopharm}+\beta_{3,\lambda}\mathit{patus}+\beta_{4,\lambda}\mathit{patgsgr}+f_{1,\lambda}(\mathit{year})+f_{2,\lambda}(\mathit{ncountry})\\&\quad+ f_{3,\lambda}(\mathit{nclaims})\\
\eta_{\pi} &=\beta_{0,\pi}+ \beta_{1,\pi}\mathit{opp}+\beta_{2,\pi}\mathit{biopharm}+\beta_{3,\lambda}\mathit{patus}+\beta_{4,\lambda}\mathit{patgsgr}+f_{1,\pi}(\mathit{year})+f_{2,\pi}(\mathit{ncountry}).
\end{aligned}
\end{equation*}
This model is denoted as ZIP\_DIC in the following.

We compare this model to the model ZIP\_NPBSS with predictors selected by the NBPSS prior where $r$ and $b$ were determined from $\alpha\in\lbrace 0.05,0.1\rbrace$, $c=0.1$. Table~\ref{tab:p} reports predictive log-scores (obtained from ten-fold cross validation) as well as values for the DIC and the widely applicable information criterion (WAIC). From the table, we can conclude, that the ZIP\_NPBSS model is clearly favoured in terms of the chosen criteria.
For the NBPSS model, we report posterior probabilities $\dsP(\delta|\yvec)$ in Table~\ref{tab:pdelta}. Based on the decision to include an effect if $\dsP(\delta|\yvec)\geq 0.5$ holds, the NBPSS prior  coincides with the stepwise approach of ZIP\_DIC for the effects of the continuous covariates but yields a sparser prediction specification for the effects of binary covariates.

\subsubsection{Bivariate Analysis of Undernutrition}
The \emph{Nigeria} data have been extracted from Demographic and Health Surveys (DHS, \url{https://dhsprogram.com/}) containing nationally representative information about the population's health and nutrition status in numerous developing and transition countries. Here we use data from Nigeria collected in 2013. Overall there are 23,042 observations after removing outliers and inconsistent observations from the data. We use \textit{stunting} and \textit{wasting} as the bivariate response vector, where \textit{stunting} refers to stunted growth measured as insufficient height of the child with respect to its age, while \textit{wasting} refers to insufficient weight for height. Hence \textit{stunting} is an indicator for chronic undernutrition while \textit{wasting} reflects acute undernutrition. We assume that the two indicators are jointly normally distributed with marginal means, marginal scales and correlation parameter depending on covariates. Specifically, the model equations for all predictors of the distributions are specified as
\begin{align*}
\eta_{k} =& \beta_{0,k}+\mathbf{\xvec}' \boldsymbol{\beta_{k}} + f_{1,k} ({\mathit{cage}}) + f_{2,k }({\mathit{mage}}) + f_{3,k}(\mathit{mbmi})+ f_{\mathit{spat},k} (\mathit{region}),
\end{align*}
where $\xvec_i$ contains 13 binary covariates characterising the household the child is living in as well as the child itself, see Table C.3 of the Online Appendix for a full description of variables. The three non-linear effects $f_1$ to $f_3$ of $\mathit{cage}$ (age of the child in months), $\mathit{mage}$ (age of the mother in years), $\mathit{mbmi}$ (body mass index of the mother) are decomposed into their linear and non-linear part as described in Section~\ref{subsec:NBPSS}. For the scale parameters, we used an exponential response function and for $\rho$ the response function $g(x)=x/\sqrt{(1+x^2)}$. The DIC/WAIC of the full model and model with NBPSS prior are 159,101/159,190 and 159,101/159,173, respectively and hence slightly better for the NBPSS prior model.

Figures~\ref{fig:nigeria:lin} and~\ref{fig:nigeria:nonlin} show the posterior means together with their 95\% posterior credible intervals of linear and non-linear effects for the full model (blue) and the model with NBPSS prior (red). For the  the function estimates $f_{j,k}=f_{j,k,\mathit{lin}}+f_{j,k,\mathit{nonlin}}$, Figure~\ref{fig:nigeria:nonlin} shows the corresponding non-linear part $f_{j,k,\mathit{nonlin}}$ separate from the linear part $f_{j,k,\mathit{lin}}$ in Figure~\ref{fig:nigeria:lin}, while the sum of the two components can be found in the Online Appendix~F.   We see that both models yield very similar point estimates, however the NBPSS prior results in slightly smoother estimates and more narrow credible intervals and hence more precise predictions -- as desired with an effective variable selection approach.
Spatial effects of the five distribution parameters with the NBPSS prior are visualized in Figure~\ref{fig:nigeria:spat}. While we omit the ones of the full model, tendencies are similar as for the remaining effects.

Inclusion probabilities are reported in Table~\ref{tab:prob:nigeria}. We find that the regional effect is relevant in all distribution parameters, i.e.~not only the marginal means but also the scales and the correlation between $\mathit{stunting}$ and $\mathit{wasting}$. Interestingly, chronic undernutrition measured by $\mathit{stunting}$ seems to be mostly driven by variables describing the life situation of the children. In contrast, besides the region of residence, the mother's nutritional status measured by $\mathit{mbmi}$ has a relevant effect only for acute undernutrition ($\mathit{wasting}$).

\subsubsection{Hedonic House Prices}\label{sec:hedonic}
We apply our methodology to the \emph{house prices} dataset of $n=98,354$ single family homes in Germany. The data were provided by F+B Research \& Consulting for Habitation, Real Estate and Environment Ltd, a business consultancy in Hamburg, Germany.  We consider the price per square metre in Euro as the response variable and explain the variation in prices in terms of four continuous covariates representing year of construction ($\mathit{yoc}$), expert rating ($\mathit{rating}$), plot area ($\mathit{areapl}$), living area ($\mathit{arealiv}$) and  spatial location ($\mathit{dist}$). We use district-specific averages $\overline{\mathit{yoc}}_\mathit{dist}$ and $\overline{\mathit{rating}}_\mathit{dist}$ as further covariates. We assume a Gaussian hierarchical location-scale model, where both expectation $\mu$ and log-variance $\log(\sigma^2)$ are related to the following hierarchical predictor.
\begin{itemize}
\item Level 1 (houses):
 \[
  \eta_k^{(1)} = f_{1,k}^{(1)}(\mathit{yoc}) + f_{2,k}^{(1)}(\mathit{rating}) + f_{3,k}^{(1)}(\mathit{areapl}) + f_{4,k}^{(1)}(\mathit{arealiv}) + f_{5,k}^{(1)}(\mathit{dist})
 \]
\item Level 2 (districts):
 \[
 \eta_\mathit{dist,k}^{(2)} = f_{1,k}^{(2)}(\overline{\mathit{yoc}}_\mathit{dist}) + f_{2,k}^{(2)}(\overline{\mathit{rating}}_\mathit{dist}) + f_{3,k}^{(2)}(\mathit{dist}),
  \]
\end{itemize}
where $f_{3,k}^{(2)}(\mathit{dist})$ follow Gaussian Markov random fields for $k=1,2$ and, as before, we decompose the effects of the continuous covariates in both levels into their linear and non-linear part such that we end up with 26 effects in total.
The NBPSS prior is put on all effects and inclusion probabilities are given in Table~\ref{tab:immo}, while Figures~\ref{fig:immo-lin} to~\ref{fig:immo-nonlin2} show the estimated  linear and non-linear parts of each function $f_{j,k}^{(l)}$, $l=1,2$ with the NBPSS prior compared to the ones of the full model. The recomposed function estimates $f_{j,k}^{(l)}=f_{j,k,\mathit{lin}}^{(l)}+f_{j,k,\mathit{nonlin}}^{(l)}$
and the estimated spatial effects can be found in the Online Appendix~G. In summary, we find that the NBPSS prior demonstrates its effect selection and shrinkage abilities also in hierarchical settings. While on level 1 the full model and the model with NBPSS prior mostly coincide, we see considerable regularisation of some non-linear effects for level 2.
The NBPSS prior is clearly able to select the spatial effect and non-linear part of  $\overline{\mathit{rating}}_\mathit{dist}$ in both distribution parameters, while the linear part and the effect of $\overline{\mathit{yoc}}_\mathit{dist}$ would be excluded according to the inclusion probabilities.


\section{Summary and Discussion}\label{sec:SumDis}
In this paper, we have developed a novel prior structure for Bayesian effect selection in structured additive distributional regression models thus extending existing approaches in terms of both flexibility of available response distributions and predictor flexibility. We derived shrinkage properties of the NBPSS prior and show its favourable properties. In simulations we demonstrate empirically that the NBPSS prior is applicable even to the selection of high dimensional coefficient blocks in more than one distribution parameter. The method promises wide applicability which we illustrate along three different examples including zero-inflated count data, a bivariate Gaussian model and a hierarchical location-scale specification for hedonic housing priors.

Instead of arbitrarily fixing hyperparameters of the inverse gamma priors we provide an intuitive and interpretable way for hyperprior elicitation which is easily accessible by applied users. This is an important feature since results react sensitively with respect to the actual choices of hyperparameters.
Yet, the NBPSS prior controls the flexibility of each effect separately since priors are assumed to be independent and does not allow to control the overall complexity of the predictor. However, the NBPSS prior could be
extended to achieve also global shrinkage properties, e.g.~by specifying the  scale parameter in the prior on $\tau^2$  as a product of a global and a local parameter~\citep{PolSco2010}. As in distributional regression the propriety of the posterior is not trivial, however, care has to be taken with respect to the specific prior choices~\citep{GhoLiMit2018}. Alternatively, if interest is rather in smoothing  and shrinkage than in explicit effect selection shrinkage priors like the double gamma prior~\cite{BitFru2016} or penalised complexity priors~\cite{SimRueMarRieSor2017} might be used.

Also, it is conceptually straightforward to include Bayesian quantile or expectile regression models into the NBPSS prior framework and we aim to do so in a future work.


\footnotesize
\renewcommand{\baselinestretch}{0.75}
\bibliography{litliste}

\newpage

\normalsize
\renewcommand{\baselinestretch}{1.2}

\input{tabs.tex}
\input{figs.tex}

\end{document}

%% file: defs.tex
\MakeShortVerb{\°}

\def \dsP {\text{$\mathds{P}$}}

\DeclareMathOperator{\SSE}{SSE}

\DeclareMathOperator{\rank}{rk}

\DeclareMathOperator{\pen}{pen}
\DeclareMathOperator{\unpen}{unpen}

\DeclareMathOperator{\spa}{span}
\DeclareMathOperator{\BetaD}{Beta}

\DeclareMathOperator{\ND}{N}

\DeclareMathOperator{\GaD}{Ga}

\DeclareMathOperator{\IGD}{IG}

\DeclareMathOperator{\BeD}{Be}
\DeclareMathOperator{\BiD}{Bi}

\def \calD {\mathcal D}

    \def \mA {\text{\boldmath$A$}}
\def \bvec {\text{\boldmath$b$}}    \def \mB {\text{\boldmath$B$}}

\def \fvec {\text{\boldmath$f$}}

    \def \mI {\text{\boldmath$I$}}
    
    \def \mK {\text{\boldmath$K$}}
    
    \def \mM {\text{\boldmath$M$}}

    \def \mP {\text{\boldmath$P$}}

    \def \mU {\text{\boldmath$U$}}
    \def \mV {\text{\boldmath$V$}}
    \def \mW {\text{\boldmath$W$}}
\def \xvec {\text{\boldmath$x$}}    \def \mX {\text{\boldmath$X$}}
\def \yvec {\text{\boldmath$y$}}    
    \def \mZ {\text{\boldmath$Z$}}
\def \nuvec {\text{\boldmath$\nu$}}

    \def \mtildeU {\text{\boldmath$\tilde U$}}
    \def \mtildeV {\text{\boldmath$\tilde V$}}

\def \ytildevec {\text{\boldmath$\tilde y$}}

\def \betavec         {\text{\boldmath$\beta$}}

\def \varepsilonvec   {\text{\boldmath$\varepsilon$}}

\def \etavec          {\text{\boldmath$\eta$}}

\def \varthetavec     {\text{\boldmath$\vartheta$}}

\def \muvec           {\text{\boldmath$\mu$}}
\def \nuvec           {\text{\boldmath$\nu$}}

\def \betatildevec         {\text{\boldmath$\tilde \beta$}}

\def \etatildevec          {\text{\boldmath$\tilde \eta$}}

\def \nullvec {\mathbf{0}}

%% file: alm.tex
\usepackage{color}
\usepackage{colordvi}
\newlength{\breite}
\breite\textwidth
\newcounter{aufg}[section]
  {\refstepcounter{aufg}\noindent\textbf{Exercise \arabic{aufg}:}
   \\*[1ex]\noindent}{\vspace{.5cm}}
   
 \newcounter{notes}[section]
  {\refstepcounter{aufg}\noindent\textbf{}
   \\*[1ex]\noindent}{\vspace{.5cm}}
   
\usepackage{amsthm}  
\newtheorem{defin}{Definition} 

\newtheorem{theo}[defin]{Theorem}

\theoremstyle{definition}
\newtheorem{remark}[]{Remark}

\newtheorem*{beisp*}{Example}
\newtheorem{Proof}{Proof}


\newtheoremstyle{break}
  {}
  {}
  {}
  {}
  {\bfseries}
  {.}
  {\newline}
  {}
  
\theoremstyle{break}

\newcommand{\head}[2]%
 {\hrule \vspace{.15cm} {\sfbold Advanced Statistical Inference, Summer Term 2012, Georg-August-University G\"ottingen}\hfill
{\sfbold Sheet #1}\\
{\sfbold Prof. Dr. Thomas Kneib, Nadja Klein}\hfill {\sfbold #2}

\vspace{.2cm}
\hrule

\vspace{1cm}

}

\newcounter{auf}
{\refstepcounter{auf}
\begin{center}
\fcolorbox[gray]{0}{.95}{
\makebox[\breite]{
\textbf{Exercise \arabic{auf}}
}}\\*[1ex]\noindent
\end{center}
}{\vspace{.5cm}}

\newcounter{loes}[section]
{\stepcounter{loes}
\begin{center}
\fcolorbox[gray]{0}{.95}{
\makebox[\breite]{
\textbf{L"osung \arabic{loes}}
}}\\*[1ex]\noindent
\end{center}
}{}

{\begin{center}
\fcolorbox[gray]{0}{.95}{
\makebox[\breite]{
\textbf{Zu Aufgabe #1}
}}\\*[1ex]\noindent
\end{center}\vspace{1cm}
}{\vspace{1cm}}

\newcounter{ka}
{\refstepcounter{ka}
\begin{center}
\framebox[\textwidth]{
\textbf{Aufgabe \arabic{ka}} \hfill #1 Punkte
}\\*[1ex]\noindent
\end{center}
}{\vspace{1cm}}

\newcounter{lka}
{\refstepcounter{lka}
\begin{center}
\framebox[\textwidth]{
\textbf{L\"osung \arabic{lka}} \hfill #1 Punkte
}\\*[1ex]\noindent
\end{center}
}{\vspace{1cm}}


%% file: tabs.tex
\begin{table}[htbp]
\begin{center}\footnotesize
\begin{tabular}{l| cccc}
\hline\hline
Data set&
  sample size & no.~of effects& distribution & computing time \\ \hline
  \emph{Patents} & 4,805 & 22 & zero-inflated Poisson & 0.25 min \\
  \emph{Nigeria} &23,042 & 108 & bivariate normal & 5.92 min \\
  \emph{House prices} & 98,354 & 26 & Gaussian location-scale & 3.75 min\\
    \hline\hline
\end{tabular}
\end{center}\caption{\footnotesize Summaries for the data sets \emph{Patents}, \emph{Nigeria}, and \emph{House prices}. Columns 2 to 4 show the number of observations, number of potential effects in the full model and the distribution for the response. The last column reports the computing time required for estimating 1,000 subsequent MCMC sweeps with the NBPSS prior.}\label{tab:summaries}
\end{table}

\begin{table}[htbp]
\begin{center}\footnotesize
\begin{tabular}{l| c  c  c  c c}
\hline\hline
Model&
  Quadratic score & Log score & Spherical score &DIC &WAIC\\ \hline
  ZIP\_DIC & -3,465.6 & -8,866.8 & 2,500.4
   & 17,136.3  & 17,214.4   \\
  ZIP\_NPBSS($\alpha=0.1$) & \textbf{-3460.1} & -8817.6 & \textbf{2511.9} & 17,124    & 17,206  \\
  ZIP\_NBPSS($\alpha=0.05$) & -3467.2 &\textbf{-8803.8} & 2507.1 &  \textbf{17,118.2} & \textbf{17,205.1} \\
    \hline\hline
\end{tabular}
\end{center}\caption{\footnotesize \emph{Patent citations}: Summarised scores in the models under consideration. Values for the predictive scores were obtained from ten-fold cross validation while DIC/WAIC are based on estimates obtained with the complete data set. The best model according to each of the criteria is highlighted in bold.}\label{tab:p}
\end{table}

\begin{table}[htbp]
\begin{center}\footnotesize
\begin{tabular}{c c| c  c| c c}
\hline\hline
\multicolumn{1}{c}{Covariate} & \multicolumn{1}{c|}{Scale}  & \multicolumn{2}{c|}{NBPSS}  & \multicolumn{2}{c}{ZIP\_DIC} \\
  & &  $\lambda$ & $\pi$ & $\lambda$ & $\pi$
  \\\hline
  $\mathit{year_{lin}}$ & continuous & $0.129$ & \textbf{1} & $\emptyset$ & $\emptyset$\\
  $\mathit{year_{nonlin}}$ & continuous& \textbf{0.965} &\textbf{0.999} & $\checkmark$ & $\checkmark$ \\
  $\mathit{ncountry_{lin}}$& continuous & $0.321$ & \textbf{0.861}& $\emptyset$ & $\emptyset$\\
    $\mathit{ncountry_{nonlin}}$ & continuous& \textbf{1}& \textbf{0.936} & $\checkmark$ & $\checkmark$\\
  $\mathit{nclaims_{lin}}$ & continuous& \textbf{0.954} & $0.286$ & $\emptyset$ & $\emptyset$\\
    $\mathit{nclaims_{nonlin}}$ & continuous& \textbf{0.996} & $0.144$ & $\checkmark$ & --\\
  $\mathit{ustwin}$& binary & $0.061$ & $0.168$ & -- & --\\
  $\mathit{opp}$& binary & $0.399$ & $0.401$ & $\checkmark$ & $\checkmark$\\
  $\mathit{biopharm}$& binary & $0.293$ & $0.381$  & $\checkmark$ & $\checkmark$\\
  $\mathit{patus}$ & binary& $0.176$ & \textbf{0.789} & $\checkmark$ & $\checkmark$ \\
  $\mathit{patgsgr}$& binary & $0.153$ & \textbf{0.571} & $\checkmark$ & $\checkmark$\\
    \hline\hline
\end{tabular}
\end{center}\caption{\footnotesize \emph{Patent citations}: Effect selection. The second column indicates the scale of the variable (continuous/binary). The third and fourth column show posterior inclusion probabilities $\dsP(\delta|\yvec)$ of $\lambda$ and $\pi$ for $\alpha =0.1$ and $c=0.1$ with the NBPSS prior. Checkmarks ($\checkmark$`') in the last two columns indicate that an effect was selected in the stepwise approach of~\citet{KleKneLan2015}, while `--' denotes the non-selected effects. Since~\citet{KleKneLan2015} did not decompose nonlinear effects into linear effects and the nonlinear deviation from this linear effect, `$\emptyset$' is used for the corresponding linear parts in ZIP\_DIC.}\label{tab:pdelta}
\end{table}

\begin{table}[htbp]
\begin{center}\footnotesize
\begin{tabular}{ c c| c  c c c c}
\hline\hline
\multicolumn{1}{c}{Covariate} & \multicolumn{5}{c}{NBPSS}\\
  & &  $\mu_{\mbox{\scriptsize wasting}}$ & $\mu_{\mbox{\scriptsize stunting}}$ & $\sigma_{\mbox{\scriptsize wasting}}$ & $\sigma_{\mbox{\scriptsize stunting}}$ & $\rho$
  \\\hline
$\mathit{bicycle}$ & binary& $0.006$ & $0.091$ & $0.005$ & $0$ & $0.01$\\
$\mathit{car}$ & binary& $0.012$ & $0.498$ & $0.008$ & $0.004$  & $0.002$\\
$\mathit{cbirthborder7}$  & binary & $0.005$ & $0.108$ & $0.005$ & $0.005$ & $0.004$\\
$\mathit{cbirthborder6}$  & binary & $0.011$ & $0.171$ & $0.006$ & $0.003$ & $0.002$\\
$\mathit{cbirthborder5}$  & binary & $0.018$ & $0.426$ & $0.005$ & $0.002$ & $0.008$\\
$\mathit{cbirthborder4}$  & binary & $0.013$ & $0.418$ & $0.004$ & $0.005$ & $0.004$\\
$\mathit{cbirthborder3}$  & binary & $0.009$ & \textbf{0.569} & $0.004$ & $0.005$ & $0.003$\\
$\mathit{cbirthborder2}$  & binary & $0.024$ & \textbf{0.846} & $0.003$ & $0.003$ & $0.002$\\
$\mathit{cbirthborder1}$  & binary & $0.007$ & \textbf{0.858} & $0.004$ & $0.007$ & $0.005$\\
$\mathit{csex}$ & binary & $0.011$ & \textbf{0.529} & $0.003$ & $0.006$ & $0.001$\\
$\mathit{ctwin}$ & binary & $0.135$ & \textbf{0.952} & $0.008$ & $0.007$ & $0.002$\\
$\mathit{electricity}$ & binary& $0.006$ & $0.194$ & $0.004$ & $0.002$ & $0.004$\\
$\mathit{motorcycle}$ & binary& $0.013$ & $0.08$ & $0.005$ & $0.002$ & $0.002$\\
$\mathit{mresidence}$ & binary& $0.027$ & $0.099$ & $0.006$ & $0.003$ & $0.002$\\
$\mathit{munemployed}$ & binary& $0.003$ & $0.069$ & $0.005$ & $0.002$ & $0.002$\\
$\mathit{radio}$ & binary& $0.005$ & $0.103$ & $0.004$ & $0.004$ & $0.002$\\
$\mathit{refrigerator}$& binary & $0.001$ & $0.458$ & $0.003$ & $0.002$ & $0.013$\\
$\mathit{television}$& binary & $0.004$ & $0.261$ & $0.008$ & $0.006$ & $0.007$\\
$\mathit{cage_{lin}}$ & continuous & $0.007$ & \textbf{1} & $0.051$ & $0.012$ & $0.067$\\
$\mathit{edupartner_{lin}}$ & binary& $0.013$ &  \textbf{0.921} & $0.004$ & $0.011$ & $0.002$\\
$\mathit{mage_{lin}}$ & continuous & $0.008$ & \textbf{0.9} & $0.006$ & $0.007$ & $0.005$\\
$\mathit{mbmi_{lin}}$ & continuous & \textbf{0.951} & \textbf{0.937} & $0.019$ & $0.007$ & $0.004$\\
$\mathit{cage_{nonlin}}$ & continuous & \textbf{1} & \textbf{1} & $0.131$ &$0.209$ & $0.393$\\
$\mathit{edupartner_{nonlin}}$ & continuous & $0.073$ & $0.204$ & $0.213$ &$0.088$ & $0.069$\\
$\mathit{mage_{nonlin}}$ & continuous & $0.078$ & $0.301$ & $0.323$ & $0.055$ & $0.086$\\
$\mathit{mbmi_{nonlin}}$ & continuous & $0.304$ & $0.12$ & $0.09$ & $0.06$ & $0.095$\\
$\mathit{region}$ & spatial& \textbf{1} & \textbf{1} & \textbf{1} & \textbf{0.999} & \textbf{0.999}
            \\
    \hline\hline
\end{tabular}
\end{center}\caption{\footnotesize \emph{Nigeria}:  Posterior inclusion probabilities $\dsP(\delta|\yvec)$ are shown in columns 3 to 7 for $\mu_{\mbox{\scriptsize wasting}}$, $\mu_{\mbox{\scriptsize stunting}}$, $\sigma_{\mbox{\scriptsize wasting}}$, $\sigma_{\mbox{\scriptsize stunting}}$ and $\rho$ with $\alpha =0.1$ and $c=0.1$ for the NBPSS prior. The second column gives the scale of the variable (continuous/binary/spatial). Effects selected according to a cut off of 0.5 are highlighted in bold.}\label{tab:prob:nigeria}
\end{table}

\begin{table}[htbp]
\begin{center}\footnotesize
\begin{tabular}{l | c  c  c  c  c  c  c  c}
\hline\hline
\multicolumn{1}{l|}{}  & Covariate  & \\\hline
 Level 1 &  $\mathit{yoc_{lin}}$ & $\mathit{yoc_{nonlin}}$ & $\mathit{rating_{lin}}$ & $\mathit{rating_{nonlin}}$ & $\mathit{areapl_{lin}}$ & $\mathit{areapl_{nonlin}}$ & $\mathit{arealiv_{lin}}$ & $\mathit{arealiv_{nonlin}}$ \\
 \hline
 $\mu$ & \textbf{1.00} & \textbf{1.00} & \textbf{1.00} & \textbf{1.00} & \textbf{1.00} & \textbf{0.67} & \textbf{1.00} & \textbf{0.94}  \\
 $\sigma^2$ &\textbf{1.00} & \textbf{1.00} & \textbf{1.00} & \textbf{1.00} & \textbf{1.00} & 0.38 & \textbf{1.00} & \textbf{1.00} \\
  \hline
Level 2  & $\mathit{\overline{\mathit{yoc}}_{lin}}$ & $\mathit{\overline{\mathit{yoc}}_{nonlin}}$ & $\mathit{\overline{\mathit{rating}}_{lin}}$ &
$\mathit{\overline{\mathit{rating}}_{nonlin}}$ &
$\mathit{dist}$ \\
\hline
$\eta_{\mathit{dist},\mu}$ & 0.29 & 0.16 & 0.93 & \textbf{1.00} & \textbf{1.00} \\
$\eta_{\mathit{dist},\sigma^2}$ & 0.18 & 0.19 & 0.41 & \textbf{0.63} & \textbf{1.00} \\
    \hline\hline
\end{tabular}
\end{center}\caption{\footnotesize \emph{House prices}: Posterior inclusion probabilities $\dsP(\delta|\yvec)$ of $\mu$ and $\sigma^2$ for  $\alpha =0.1$ and $c=0.1$ (first row) and of $\eta_{\mathit{dist},\mu}$ and $\eta_{\mathit{dist},\sigma^2}$ (second row) with the NBPSS prior.}\label{tab:immo}
\end{table}

%% file: figs.tex
\begin{figure}[ht]
\begin{center}
\includegraphics[scale=0.5]{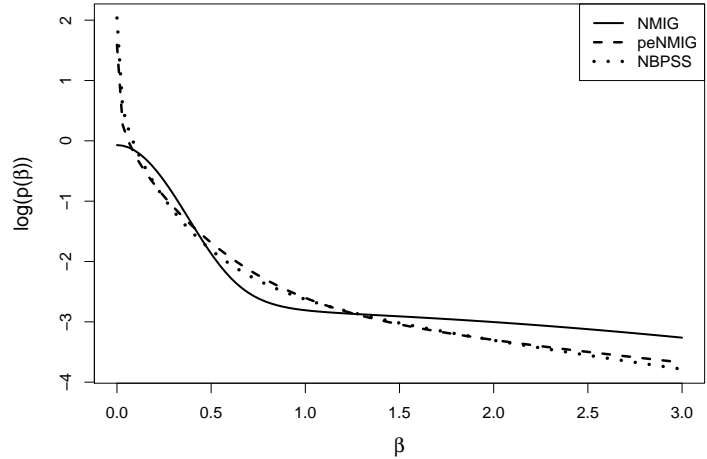}
\end{center}
\caption{\footnotesize Univariate marginal log-densities for a standard NMIG prior (solid line), the peNMIG prior of \citet[][dashed line]{schfah11} and the NBPSS prior (dotted line). Hyperparameters are set to $a_0=b_0=1$, $a=5$, $b=50$, $r=0.005$.\label{fig:marginaldensities1d}}
\end{figure}

\begin{figure}[ht]
\begin{center}
\includegraphics[width=0.95\textwidth]{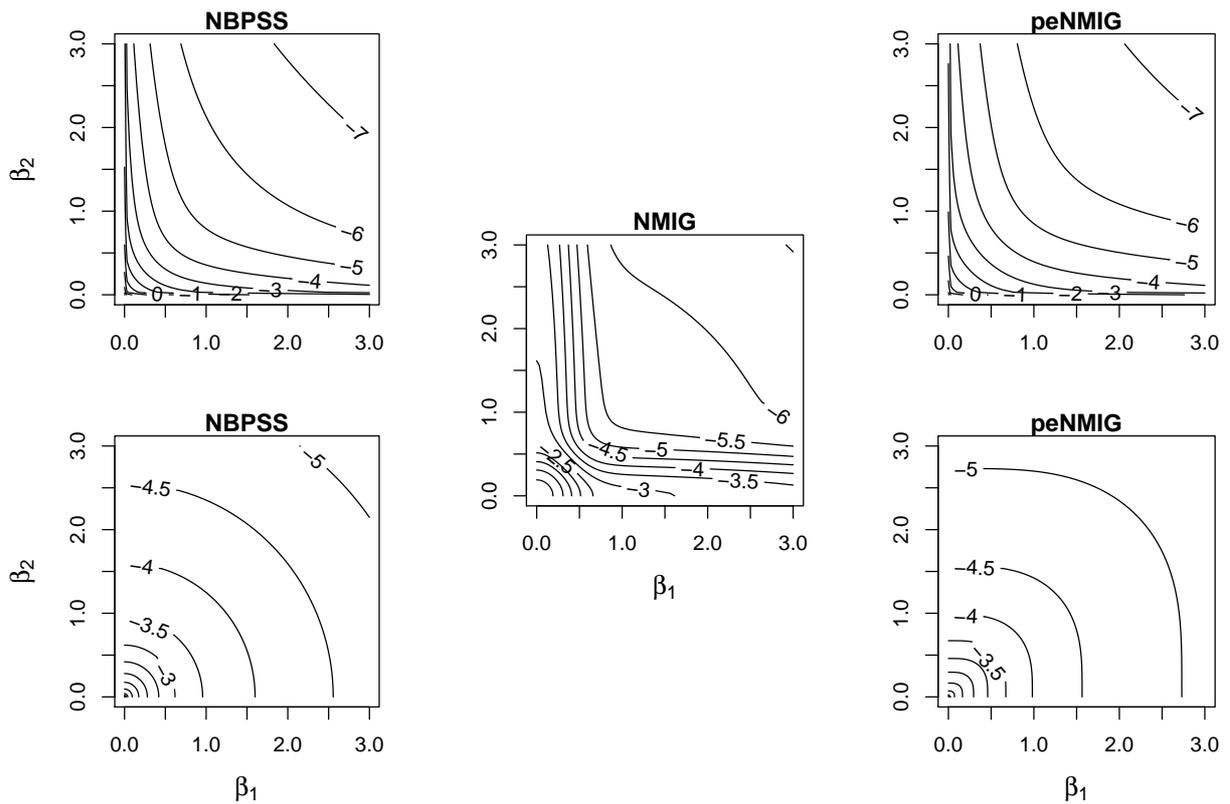}
\end{center}
\caption{\footnotesize Contour lines of bivariate marginal log-densities for a standard NMIG prior (middle panel), the peNMIG (right column) and the NBPSS prior (left column). The first row panels show results for parameters with distinct hyperparameters and the second row panels show results for parameters sharing the same $\tau$. For the standard NMIG, the hyperparameters are by construction assumed to be distinct and no changes in the row are possible.\label{fig:marginaldensities2d}}
\end{figure}

\begin{figure}[ht]
\begin{center}
\includegraphics[scale=0.5]{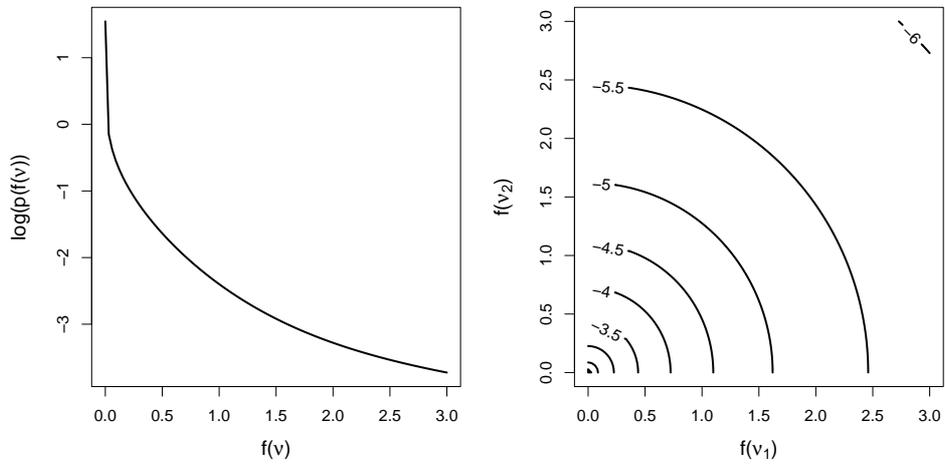}
\end{center}
\caption{\footnotesize Univariate (left) and bivariate (right) marginal log-densities of $f(\nuvec)$. The hyperparameters have been fixed at $a=5$, $b=50$, $r=0.005$ and $a_0=b_0=1$.\label{fig:marginaldensitiesf}}
\end{figure}

\begin{figure}[ht]
\begin{center}
\includegraphics[scale=0.5]{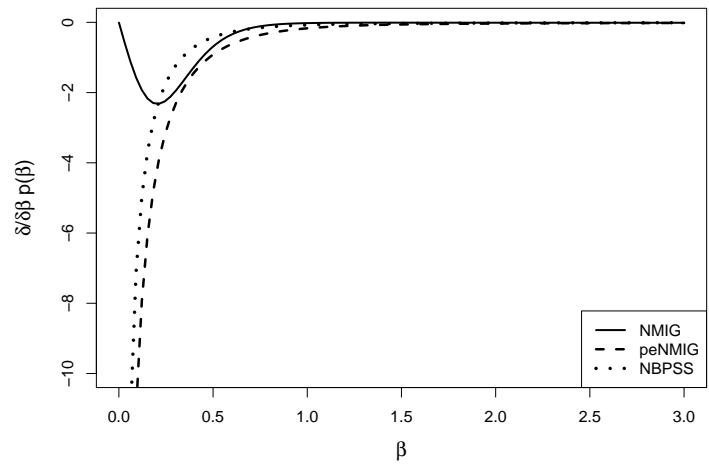}
\end{center}
\caption{\footnotesize Score function of the marginal prior $p(\betavec)$ for the standard NMIG (solid line), the parameter expanded prior by \citet[][dashed line]{schfah11} and the parameter expanded prior proposed in this paper (dotted line).\label{fig:scorefun}}
\end{figure}

\begin{figure}\hspace{-0.2cm}\flushleft{\includegraphics[width=1.00\textwidth,angle= 0]{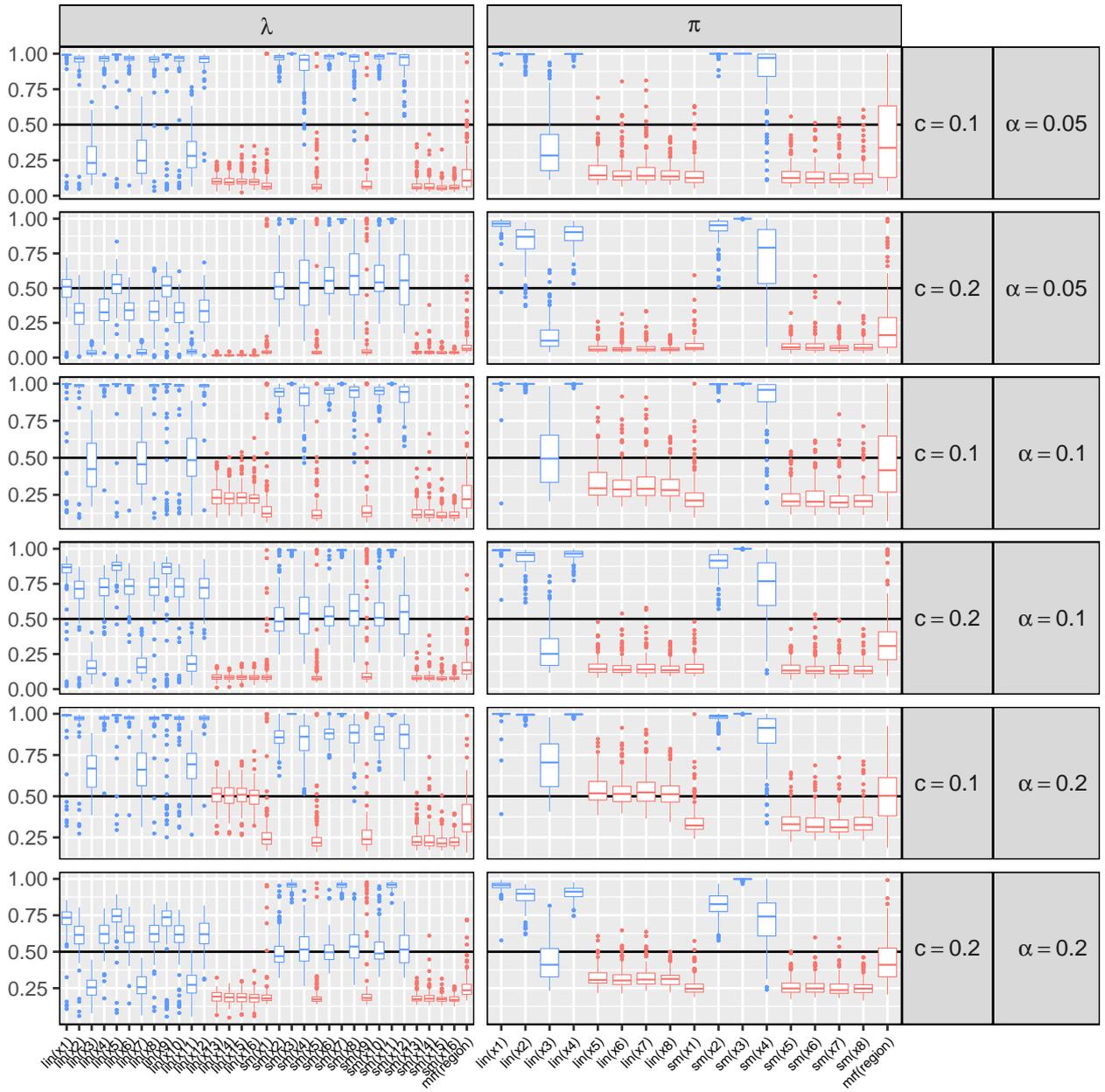}}\caption{\footnotesize Posterior inclusion probabilities of effects in the zero-inflated Poisson model with $n = 2,000$ observations, uncorrelated covariates and no true spatial effect in the predictor (i.e. the data generating model does not comprise a spatial effect but we estimate a model including a spatial effect) . Blue boxplots correspond to effects that are included in the true model while the red boxes correspond to the noise variables that do not have an effect in the data generating mechanism.}\label{fig:zip_lambda_pi}\end{figure}

\begin{figure}\centering{\includegraphics[width=0.85\textwidth,angle= 0]{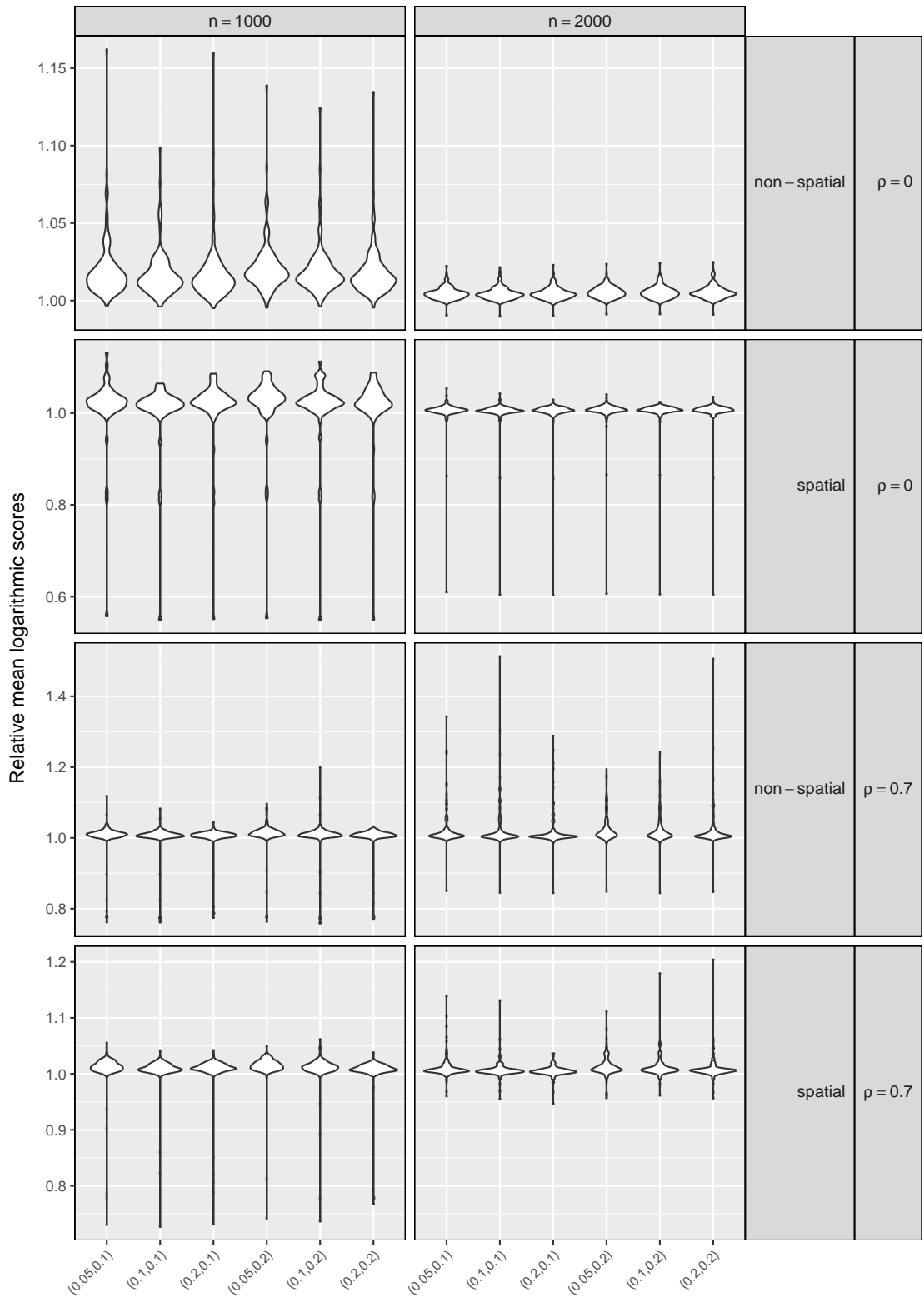}\caption{\footnotesize Violin plots of relative mean log-scores (i.e.~mean log scores obtained with the NBPSS prior divided by mean log scores of the oracle model) in the zero-inflated Poisson model. The log-scores are averaged over 5,000 new test data observations for each simulation replicate. The columns represent the different sample sizes $n=1,000;2,000$, rows 1 and 3 belong to the non-spatial scenarios (no spatial effect in the data generating model) and rows 2 and 4 to the spatial ones (the data generating model comprises a spatial effect). Covariates are uncorrelated in rows 1 and 2 and correlated in rows 3 and 4. The different boxplots within a column/row correspond to different combinations of $\alpha$, $c$ denoted as $(\alpha,c)$ in the labels.}\label{fig:logscores}
}\end{figure}

\begin{figure}\centering{\includegraphics[width=0.8\textwidth,angle=0]{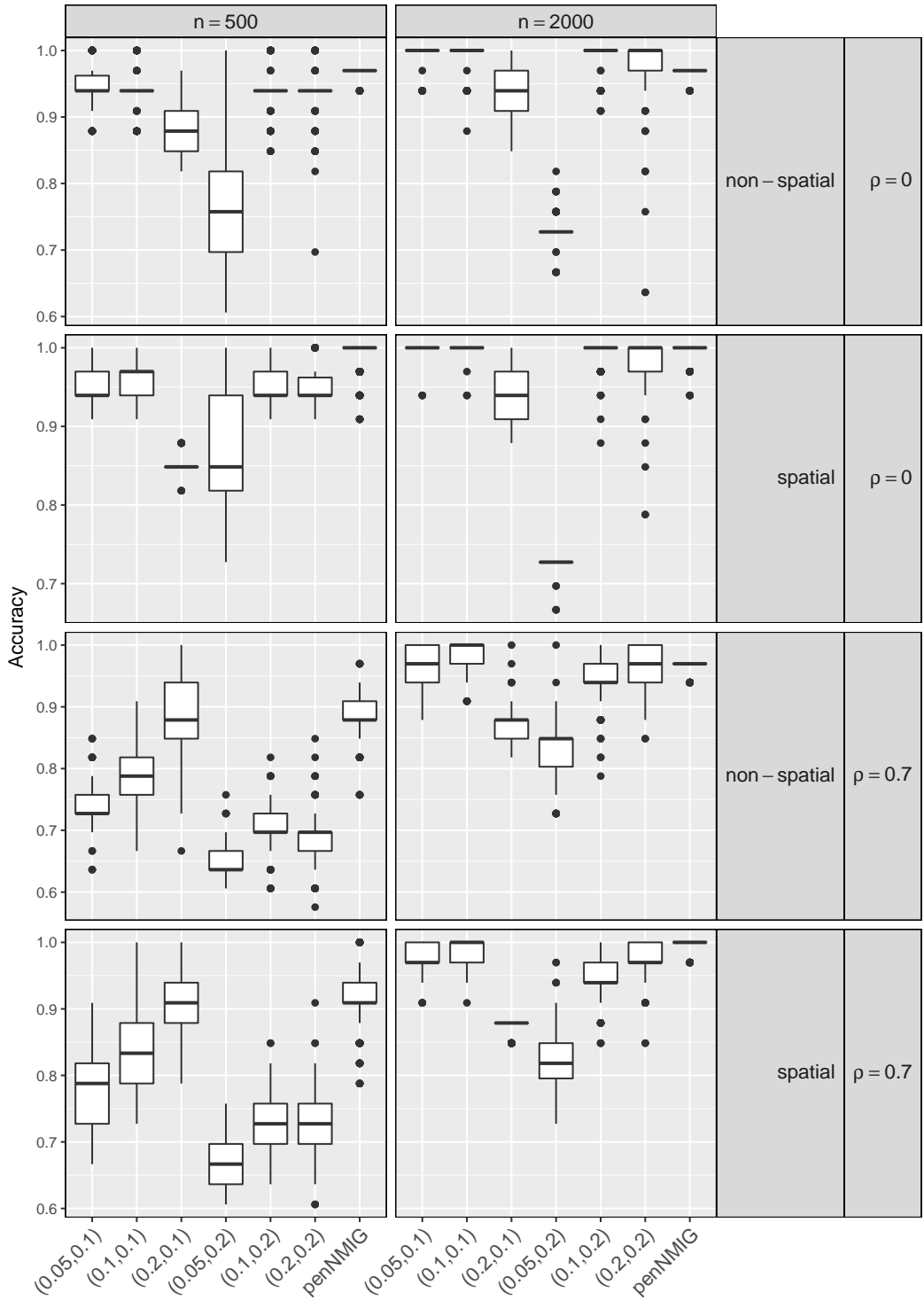}}\caption{\footnotesize Overall accuracy  (measured by the sum of true positives and true negatives divided by the total number of effects) for the Poisson model. The columns represent the different sample sizes $n=500;2,000$, rows 1 and 3 belong to the non-spatial scenarios (no spatial effect in the data generating model) and rows 2 and 4 to the spatial ones (the data generating model comprises a spatial effect). Covariates are uncorrelated in rows 1 and 2 and correlated in rows 3 and 4. Last, the boxplot on the right of each subplot shows the peNMIG prior results, the remaining ones correspond to different choices of $\alpha$ and $c$ of the NBPSS prior, denoted as $(\alpha,c)$ in the labels.}\label{fig:acc_poisson}\end{figure}

\begin{figure}\centering{\includegraphics[angle= 0]{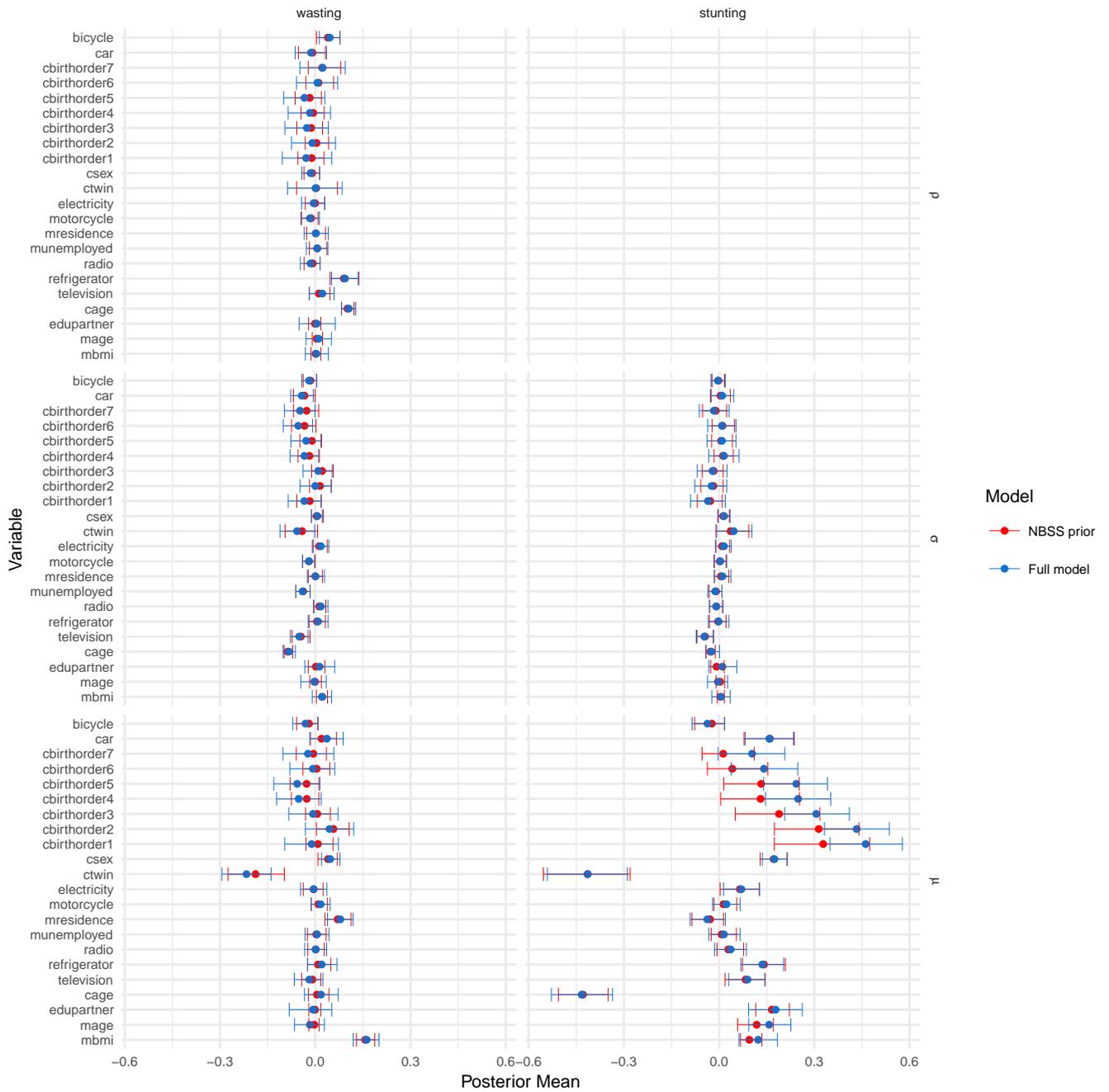}}\caption{\footnotesize Nigeria: Posterior means and 95\% credible intervals for the linear effects of all model parameters (left column for stunting, right column for wasting, top row for $\rho$, middle row for $\sigma$, bottom row for $\mu$). Since $\rho$ acts on both responses, the effects are only shown in the first column. Red corresponds to results for the NBPSS prior and blue to the full model.}\label{fig:nigeria:lin}
\end{figure}

\begin{figure}\centering{\includegraphics[angle= 0,width=0.95\textwidth]{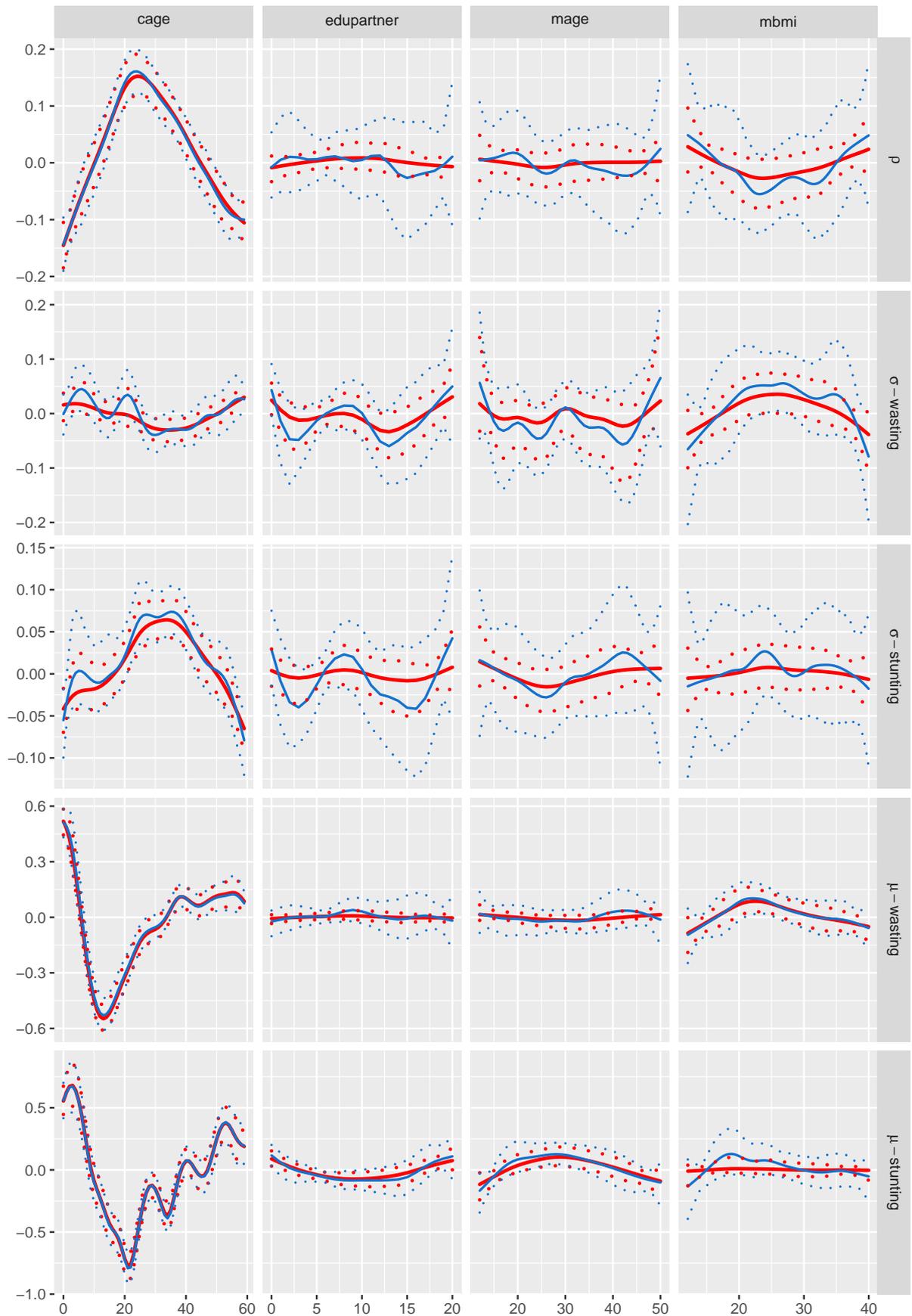}}\caption{\footnotesize \emph{Nigeria}: Posterior means and pointwise 95\% credible intervals for the non-linear effects $f_{j,k,\mathit{nonlin}}$ of $\mathit{cage},\mathit{mage}$, $\mathit{mbmi}$ and $\mathit{edupartner}$ (column-wise) for all model parameters ($\rho,\sigma_{\mbox{\scriptsize stunting}},\sigma_{\mbox{\scriptsize wasting}},\mu_{\mbox{\scriptsize stunting}},\mu_{\mbox{\scriptsize wasting}}$, row-wise). Red corresponds to results for the NBPSS prior and blue to the full model.}\label{fig:nigeria:nonlin}
\end{figure}

\begin{figure}\centering{\includegraphics[width=0.47\textwidth,angle= 0]{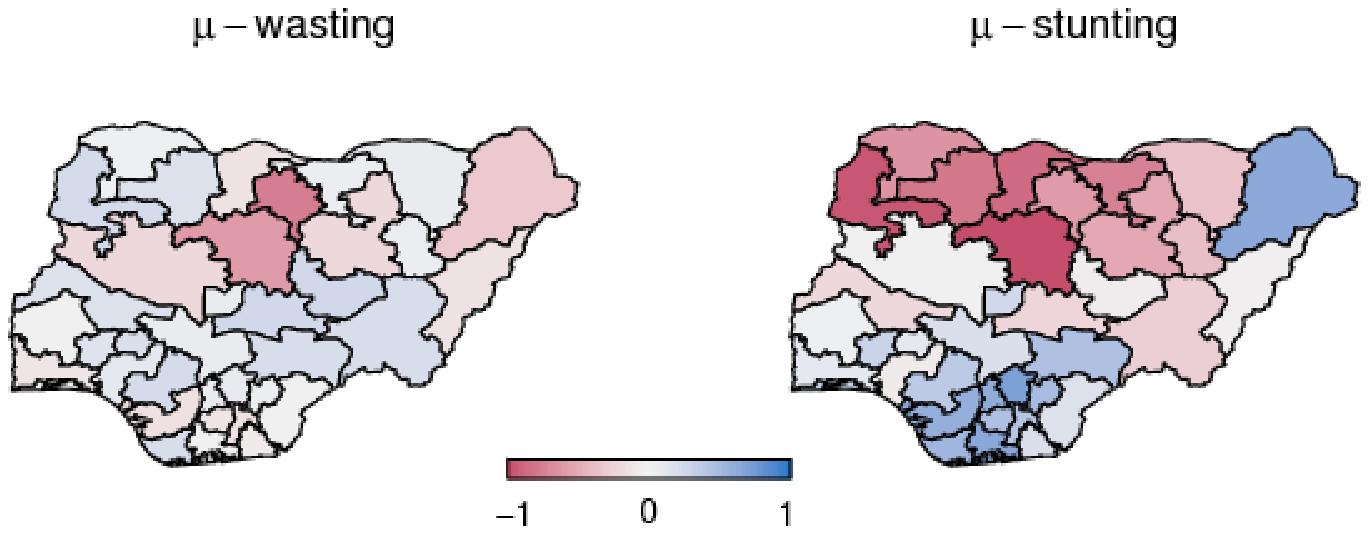}\hspace{0.9cm}
\includegraphics[width=0.47\textwidth,angle= 0]{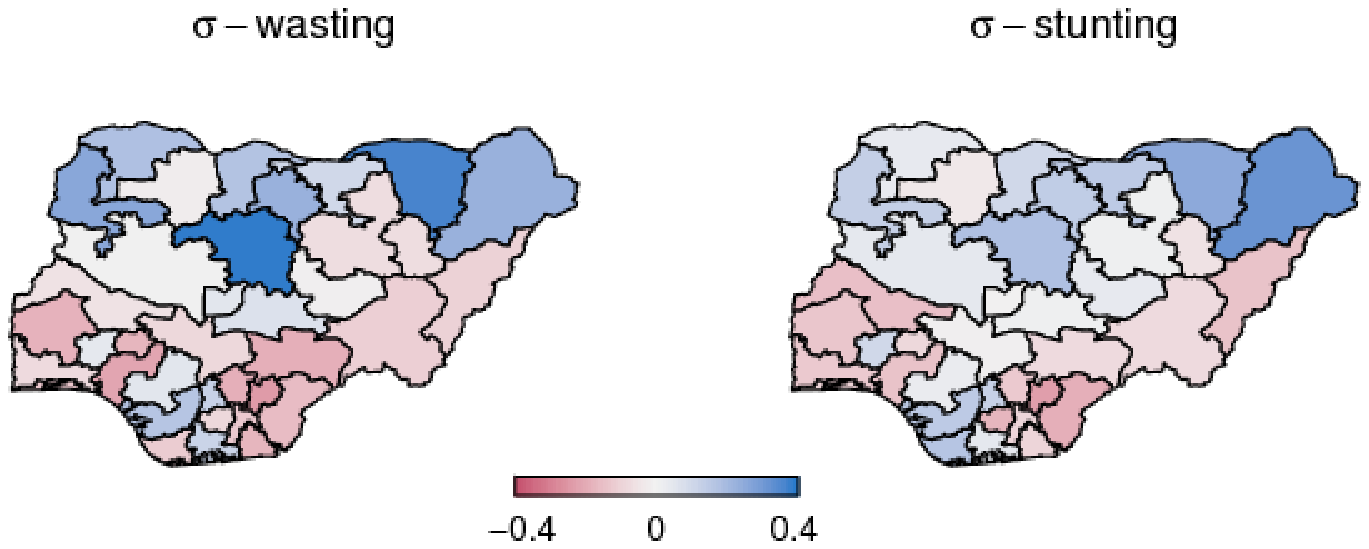}}\\
\centering{\includegraphics[width=0.20\textwidth,angle= 0]{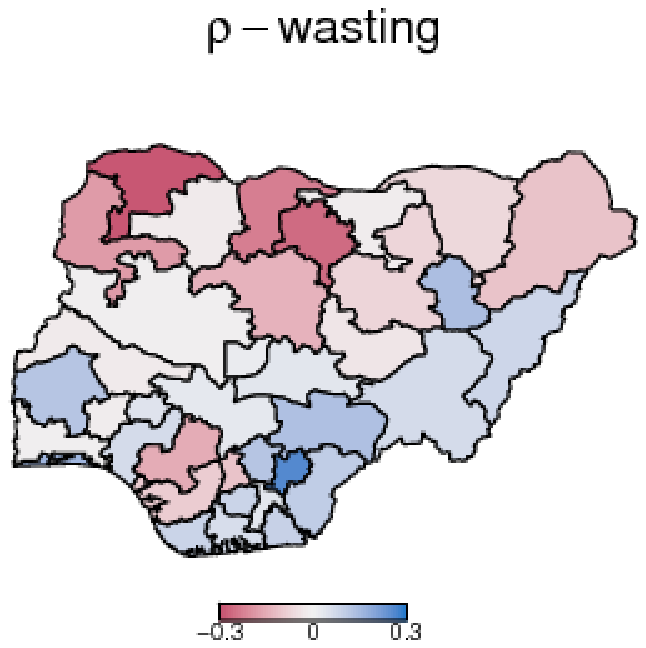}}
\caption{\footnotesize \emph{Nigeria}: Posterior means for the spatial effects of all model parameters $\mu_{\mbox{\scriptsize wasting}}$, $\mu_{\mbox{\scriptsize stunting}}$, $\sigma_{\mbox{\scriptsize wasting}}$, $\sigma_{\mbox{\scriptsize stunting}}$ estimated with the NBPSS prior.}\label{fig:nigeria:spat}
\end{figure}

\begin{figure}\centering{\includegraphics[width=0.8\textwidth,angle= 0]{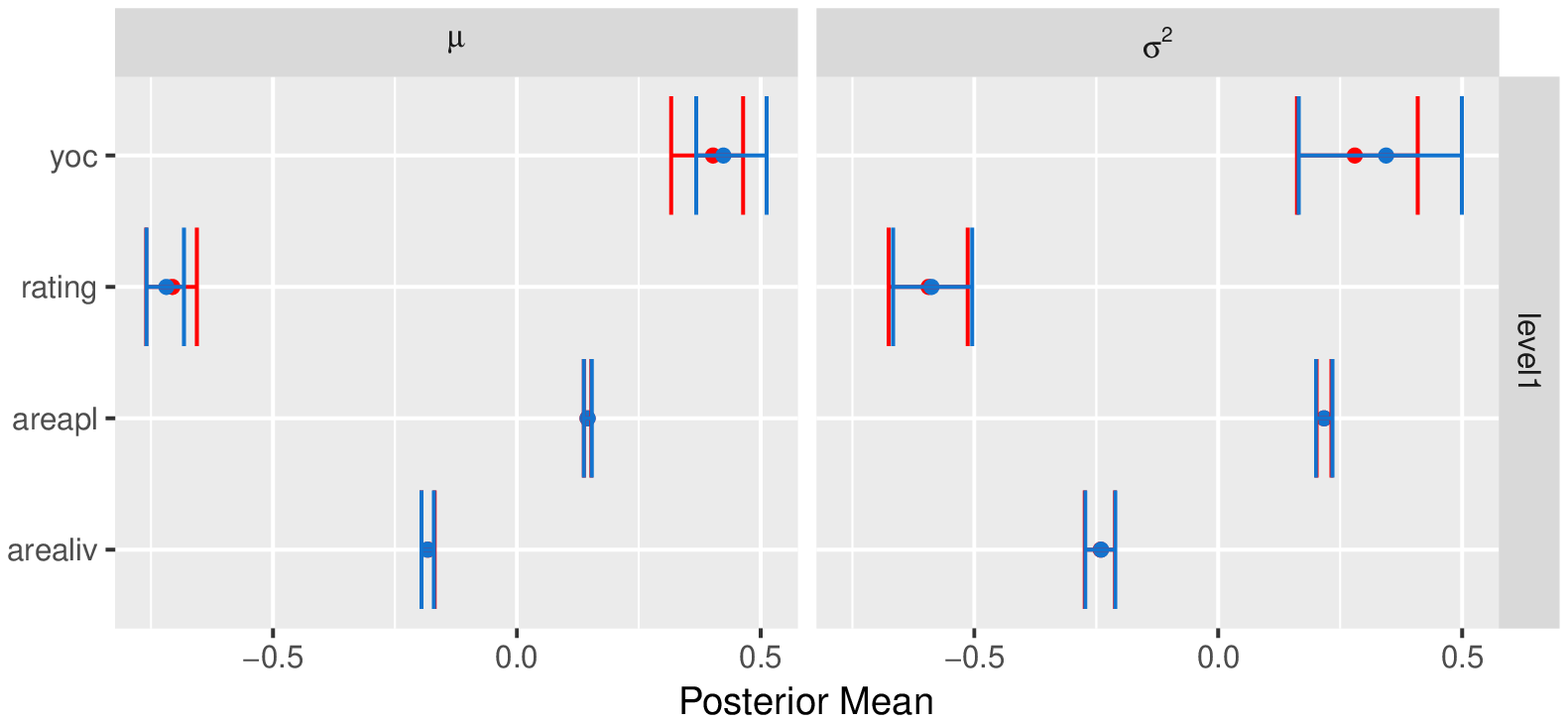}
\includegraphics[width=0.90\textwidth,angle= 0]{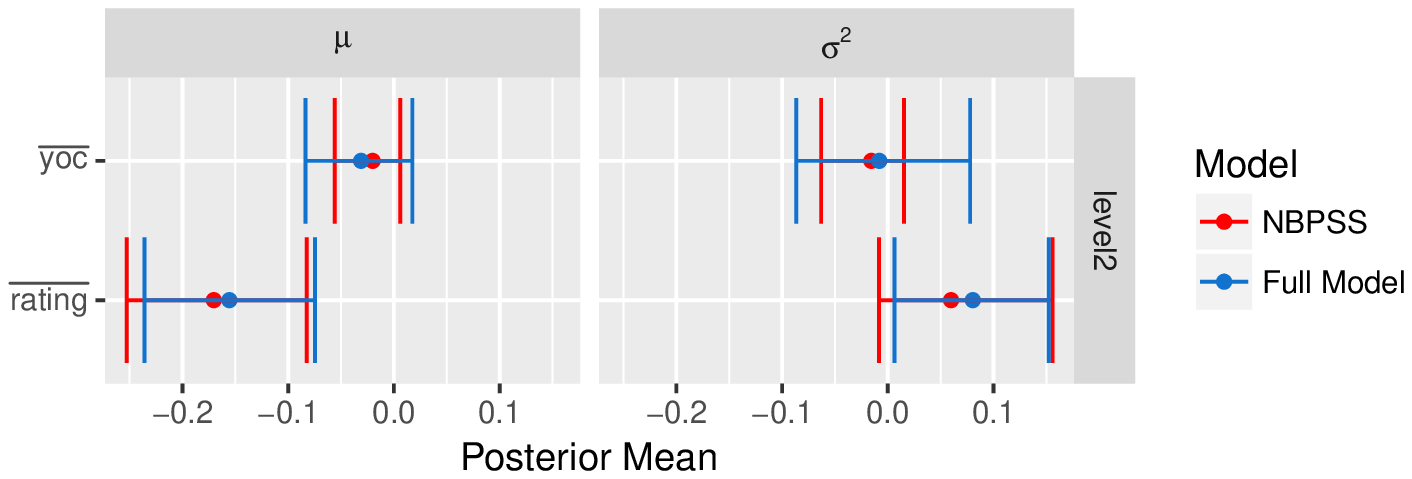}}
\caption{\footnotesize \emph{House prices}: Estimated posterior mean linear effects with 95\% credible intervals of the model parameters  $\mu$, $\sigma^2$ (level 1, first row), $\eta_\mathit{dist,\mu}$ and $\eta_\mathit{dist,\sigma^2}$ (level 2, second row) estimated with the NBPSS prior.}\label{fig:immo-lin}
\end{figure}

\begin{sidewaysfigure}\centering{\includegraphics[width=0.60\textwidth,angle= -90]{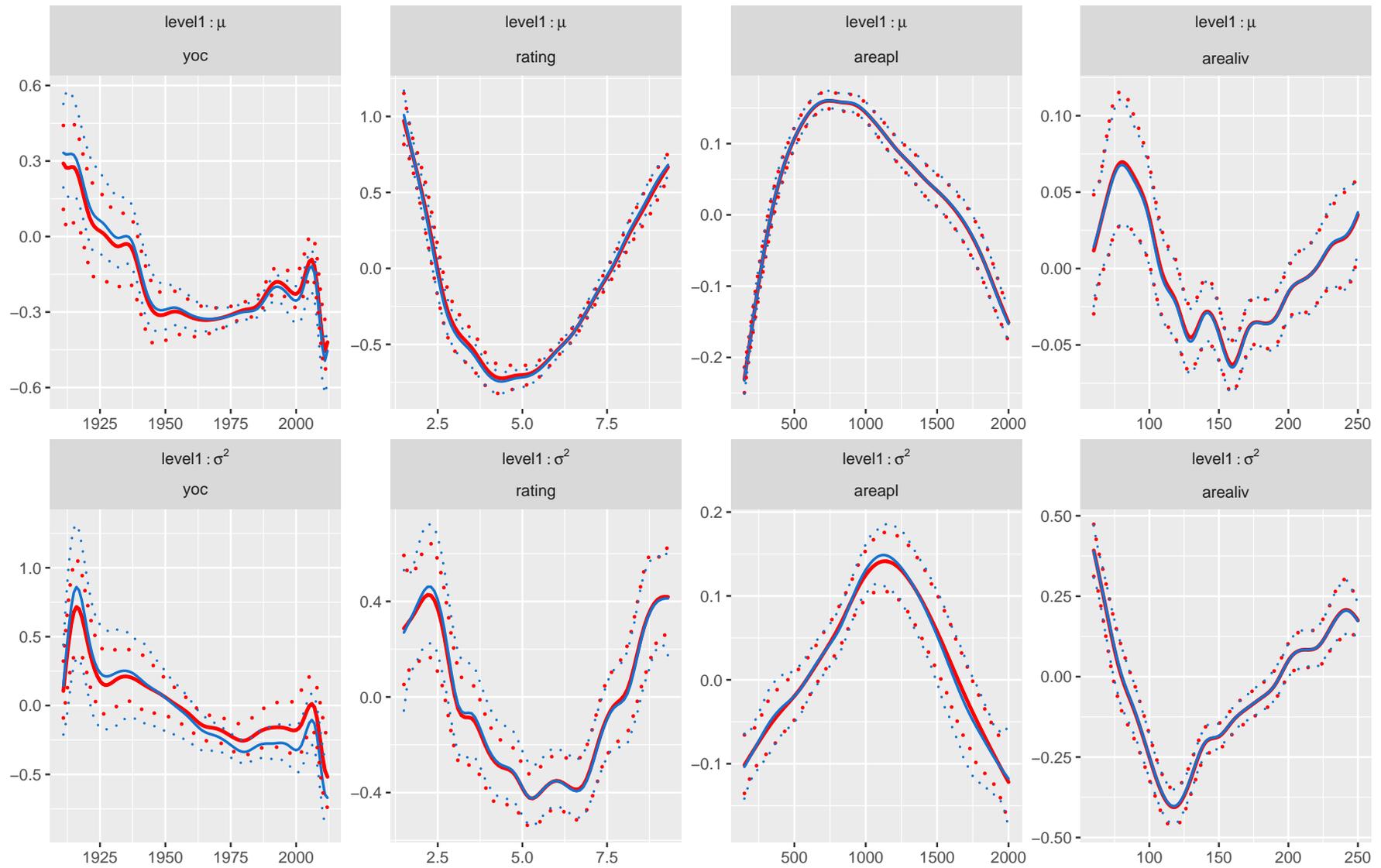}}
\caption{\footnotesize \emph{House prices}: Estimated posterior mean non-linear effects with 95\% credible intervals of the model parameters  $\mu$, $\sigma^2$ (level 1) estimated with the NBPSS prior (red) and the full model (blue).}\label{fig:immo-nonlin}
\end{sidewaysfigure}

\begin{figure}
\centering{\includegraphics[width=0.8\textwidth,angle= 0]{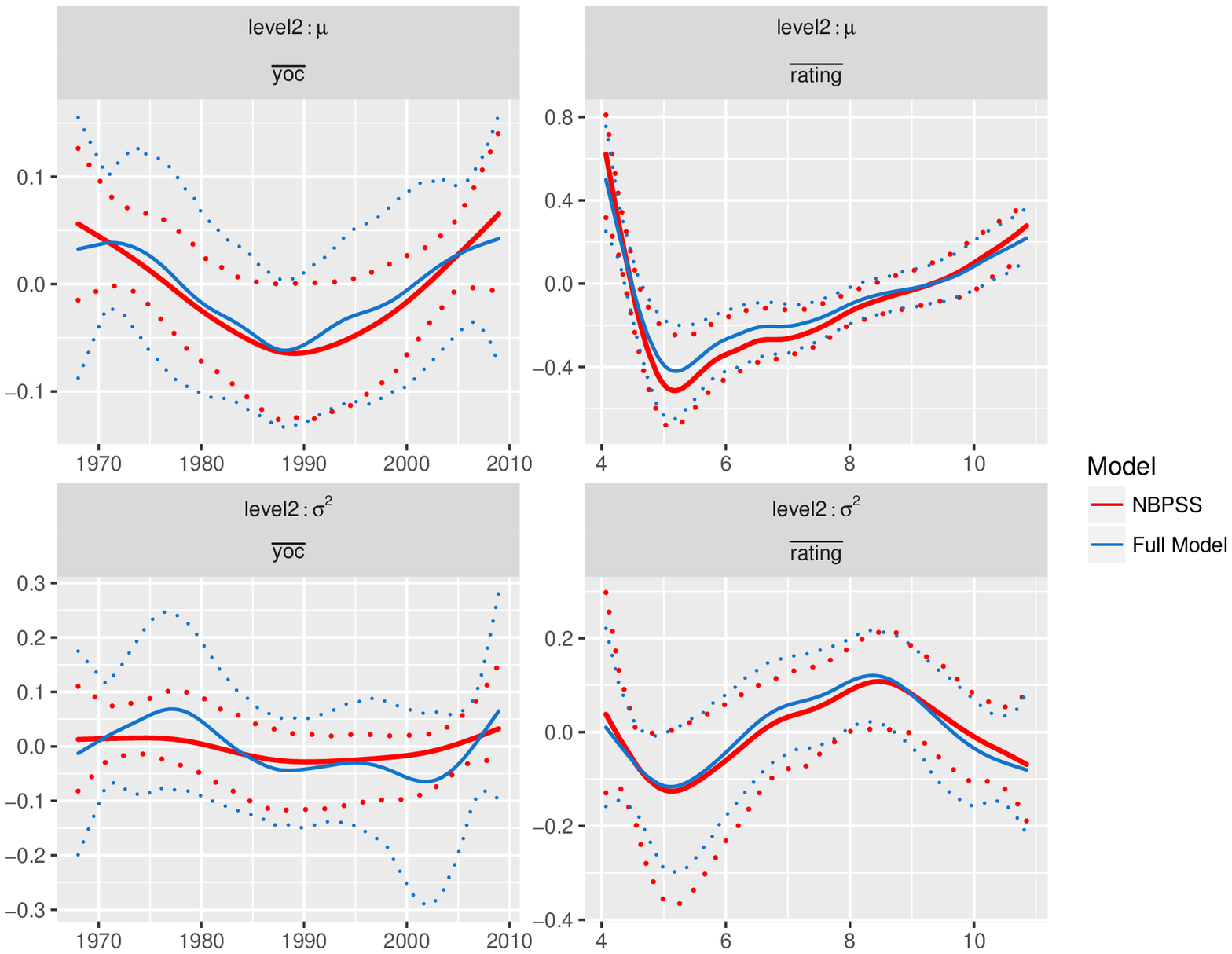}}
\caption{\footnotesize \emph{House prices}: Estimated posterior mean non-linear effects with 95\% credible intervals of the model parameters   $\eta_\mathit{dist,\mu}$ and $\eta_\mathit{dist,\sigma^2}$ (level 2) estimated with the NBPSS prior (red) and the full model (blue).}\label{fig:immo-nonlin2}
\end{figure}